%% file: 0_Main1.tex
\documentclass[pdflatex,sn-mathphys-num]{sn-jnl}% Math and 

\usepackage{graphicx}%
\usepackage{multirow}%
\usepackage{amsmath,amssymb,amsfonts}%
\usepackage{amsthm}%
\usepackage{mathrsfs}%
\usepackage[title]{appendix}%
\usepackage{xcolor}%
\usepackage{textcomp}%
\usepackage{manyfoot}%
\usepackage{booktabs}%
\usepackage{algorithm}%
\usepackage{algorithmicx}%
\usepackage{algpseudocode}%
\usepackage{listings}%
\usepackage{comment}
\newcommand{\tabincell}[2]{\begin{tabular}{@{}#1@{}}#2\end{tabular}}
\usepackage{color}

\theoremstyle{thmstyleone}%
\theoremstyle{thmstyletwo}%

\theoremstyle{thmstylethree}%

\begin{document}

\title{Domain-constrained Synthesis of Inconsistent Key Aspects in Textual Vulnerability Descriptions}

\author[1]{\fnm{Linyi} \sur{Han}}\email{hanly2@tju.edu.cn}
\equalcont{These authors contributed equally to this work.}

\author[2]{\fnm{Shidong} \sur{Pan}}\email{Shidong.Pan@data61.csiro.au}
\equalcont{These authors contributed equally to this work.}

\author[2]{\fnm{Zhenchang} \sur{Xing}}\email{Zhenchang.Xing@data61.csiro.au}

\author[1]{\fnm{Sofonias} \sur{Yitagesu}}\email{sofoniasyitagesu@yahoo.com}

\author*[1]{\fnm{Xiaowang} \sur{Zhang}}\email{xiaowangzhang@tju.edu.cn}

\author[1]{\fnm{Zhiyong} \sur{Feng}}\email{zyfeng@tju.edu.cn}

\author[2]{\fnm{Jiamou} \sur{Sun}}\email{Frank.Sun@data61.csiro.au}

\author[3]{\fnm{Qing} \sur{Huang}}\email{qh@jxnu.edu.cn}

\affil*[1]{\orgname{Tianjin University}, \state{Tianjin}, \country{China}}

\affil[2]{\orgname{CSIRO's Data61}, \city{Canberra}, \state{ACT}, \country{Australia}}

\affil[3]{\orgname{Jiangxi Normal University},  \city{Nanchang}, \state{Jiangxi}, \country{China}}

%\affil[3]{\orgname{The Australian National University}, \orgaddress{\street{Acton}, \city{Canberra}, \postcode{2601}, \state{ACT}, \country{Australia}}}

\input{1_Abstract}

\maketitle
\input{2.1_Introduction}

\input{3_FormativeStudy}
\input{4_LabelDesign}
\input{5_Approach_V1}

\input{7_Results}

\input{8_Disscution}
\input{9.1_RelatedWork}
\input{10_Conclusion}

\bibliography{11_Reference}% common bib file

\end{document}

%% file: 1_Abstract.tex
\abstract{Textual Vulnerability Descriptions (TVDs) are crucial for security analysts to understand and address software vulnerabilities. 
However, the key aspect inconsistencies in TVDs from different repositories pose challenges for achieving a comprehensive understanding of vulnerabilities. 
Existing approaches aim to mitigate inconsistencies by aligning TVDs with external knowledge bases, but they often discard valuable information and fail to synthesize comprehensive representations.
In this paper, we propose a domain-constrained LLM-based synthesis framework for unifying key aspects of TVDs. Our framework consists of three stages: 1) Extraction, guided by rule-based templates to ensure all critical details are captured; 2) Self-evaluation, using domain-specific anchor words to assess semantic variability across sources; and 3) Fusion, leveraging information entropy to reconcile inconsistencies and prioritize relevant details.
This framework improves synthesis performance, increasing the F1 score for key aspect augmentation from 0.82 to 0.87, while enhancing comprehension and efficiency by over 30\%. We further develop Digest Labels, a practical tool for visualizing TVDs, which human evaluations show significantly boosts usability.}

\keywords{Vulnerability Analysis, Inconsistent Key Aspect, Domain-Constrained Synthesis, Digest Labels System, Vulnerability Repositories}

%% file: 2.1_Introduction.tex
\section{Introduction}
\textcolor{black}{
Software vulnerabilities pose severe risks to modern systems, as they can be exploited by attackers to compromise confidentiality, integrity, and availability. 
To facilitate mitigation, various vulnerability databases have been established, such as the Common Vulnerabilities and Exposures (CVE)~\cite{CVE}, National Vulnerability Database (NVD)~\cite{nistnvd}, and IBM X-Force~\cite{ibm}. 
These repositories provide \textit{Textual Vulnerability Descriptions} (TVDs), which contain crucial information for security analysts to understand, assess, and repair vulnerabilities. 
TVDs are expected to capture essential \textbf{key aspects} of a vulnerability, including the \textit{Vulnerability Type}, how an attack can be executed (\textit{Attack Vector}), the type of attacker (\textit{Attacker Type}), the \textit{Root Cause}, and its potential \textit{Impact}. 
}

\begin{figure}[tbp]
    \centering
    \includegraphics[width=0.85\textwidth]{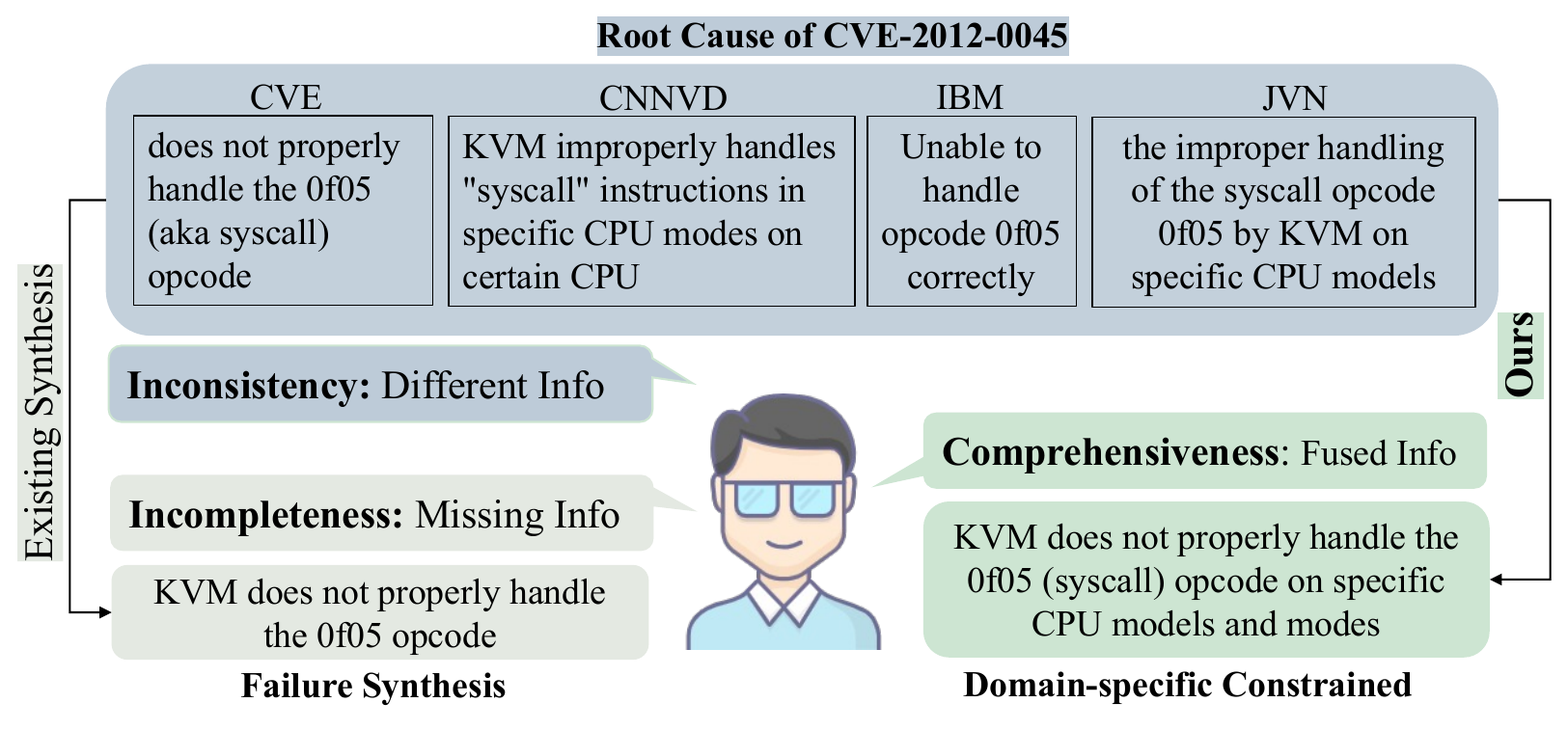}
    \caption{Different repositories provide varying details. If synthesis loses details, the purpose of achieving comprehensive vulnerability understanding is undermined.}
    \label{intro}
\end{figure}

\textcolor{black}{
However, inconsistencies often exist between different repositories for the same vulnerability. 
These inconsistencies may arise because one TVD provides only a partial description, while another emphasizes specific contextual details. 
For instance, as illustrated in Figure~\ref{intro}, the \textit{Root Cause} of CVE-2012-0045 is described in different ways: CVE provides a basic description (``does not properly handle the 0f05 opcode''), whereas CNNVD offers more detailed information (``KVM improperly handles syscall instructions in specific CPU modes on certain CPU models''). 
Such differences reflect not errors but multiple perspectives of the same issue. 
Drawing from the \textit{Blind Men and an Elephant} theory, we argue that instead of discarding inconsistent information, synergistically integrating these complementary perspectives leads to a more comprehensive understanding of vulnerabilities.
}

\textcolor{black}{
These inconsistencies present a significant challenge for analysts. 
Given the large volume of vulnerabilities and the technical complexity of TVDs, analysts must spend additional effort to reconcile descriptions from multiple repositories. 
Manually synthesizing this information is time-consuming and error-prone. 
Recent studies attempt to address the issue by comparing TVDs with external knowledge bases to filter out inconsistencies. 
Yet, this approach often treats differences as errors and discards details that may in fact provide valuable complementary information. 
Consequently, important contextual details are lost, which hinders the ultimate goal of comprehensive vulnerability understanding.
}

\textcolor{black}{
Large Language Models (LLM) offer promising opportunities for synthesizing TVDs, as they are capable of semantic reasoning and integrating diverse inputs. 
However, existing LLM-based synthesis methods are primarily guided by soft semantic instructions, such as ``synthesize vulnerability descriptions with as much detail as possible''. 
These instructions provide only high-level direction and lack the precision to enforce domain-specific boundaries. 
As a result, generated descriptions may remain vague and overlook critical information. 
For example, as shown in Figure~\ref{intro}, when synthesizing the \textit{Root Cause} of CVE-2012-0045, an unconstrained LLM may output ``KVM fails to handle the 0f05 syscall opcode'', missing key details such as ``specific CPU models'' or ``modes''.
}

\textcolor{black}{
To overcome these challenges, this paper proposes a domain-constrained LLM-based synthesis approach for TVDs. 
By introducing domain-specific conditions that restrict the model’s searching space, our approach preserves diverse expressions while ensuring that critical technical details are retained. 
Based on this framework, we further develop a practical tool, named \textbf{Digest Labels (DLs)}, which standardizes the synthesis of TVDs from multiple repositories. 
In real-world evaluations, DLs significantly improve analysts’ comprehension and efficiency, achieving a 31.3\% increase in comprehensiveness and a 33.5\% increase in efficiency. 
}

\textcolor{black}{
Overall, this work makes the following contributions:
\begin{itemize}
\item \textbf{Problem Identification:} We reveal that inconsistencies across vulnerability repositories are widespread and stem from complementary perspectives rather than simple errors. 
\item \textbf{Domain-Constrained Synthesis:} We introduce a novel synthesis approach that integrates LLM with domain-specific constraints to preserve completeness and accuracy of TVDs. 
\item \textbf{Digest Labels (DLs):} We design and implement the first nutrition-label inspired system for TVDs, demonstrating its usability and effectiveness through empirical experiments and human evaluation. 
\end{itemize}
}

\textcolor{black}{
The remainder of this paper is organized as follows: 
Section~\ref{formative_s} provides the motivation for synthesizing key aspects. 
Section~\ref{labelDesign} introduces the design of Digest Labels (DLs) and explains how labels are structured to represent synthesized information. 
Section~\ref{approach} then presents our domain-constrained synthesis approach. 
Section~\ref{result} reports the experimental evaluation, while Section~\ref{discussion} discusses implications and limitations. 
Section~\ref{related works} reviews related work, and Section~\ref{Conclusion} concludes the paper.
}

%% file: 3_FormativeStudy.tex
\section{Motivation for Synthesis of Key Aspect in TVDs}
\label{formative_s}
In this motivation study, we systematically retrospect the challenges associated with CVE TVD from three perspectives.

\subsection{Industry Standards}

Vulnerability management is one of the constitutional problem of software engineering industry. 
Many authoritative organizations launch a set of industry standards of CVE, including the TVD standardization, with the goal of improving overall quality.
Those standards provide the best of practices to all practitioners as reference and we highlight the content related to TVD standardization:
\begin{itemize}

    \item \textbf{[NIST]} The National Institute of Standards and Technology (NIST), under the U.S. Department of Commerce, offers essential standards for vulnerability management. These include naming schemas (NIST SP 800-51), formalization (NIST SP 800-231), and communication guidelines (NIST SP 800-216). The Security Content Automation Protocol (SCAP) provides a standardized way to communicate vulnerability information. As stated by CISA: \textit{``a suite of specifications for standardizing the format and nomenclature by which software flaw and security configuration information is communicated, both to machines and humans''}~\cite{cisa2022vulnerability}.

    \item \textbf{[ISO]} The International Organization for Standardization (ISO) released BS ISO/IEC 29147:2018, which provides a standard format for reporting software name and version of vulnerabilities. However, no formal guidelines exist for key aspects in TVDs, making them harder to standardize due to their human-originated nature. As stated in the standard, \textit{``a vulnerability is generally a behaviour or set of conditions that allows the violation of an explicit...''}~\cite{iso29147}. Our work enhances these specifications by addressing the gaps in formatting key aspects.
\end{itemize}
Even though these standards set expectations for online CVE TVDs, we have observed prevalent issues of missing information and inconsistency, specifically discussed in the following section.

\textcolor{black}{
\subsection{Definition of Key Aspects}
\label{sec:key_aspects}
We clarify what we mean by \textit{key aspects} of a TVD.  
Drawing from prior work~\cite{DBLP:journals/tosem/GuoCXLBS22, DBLP:conf/uss/DongGCXZ019, DBLP:journals/tdsc/HeWZWZLY24, DBLP:journals/compsec/SunOZLWZ23} and industry standards such as CVE~\cite{CVE} and IBM X-Force~\cite{ibm}, we define the following key aspects:}

\textcolor{black}{
\begin{itemize}
    \item \textbf{Vulnerability Type:} the category or nature of the vulnerability (e.g., buffer overflow, SQL injection).  
    \item \textbf{Attack Vector:} how an attack can be executed (e.g., remote network access, local execution).  
    \item \textbf{Attacker Type:} the assumed capabilities or privileges of the attacker (e.g., authenticated user, remote adversary).  
    \item \textbf{Root Cause:} the underlying technical reason leading to the vulnerability (e.g., improper opcode handling, incorrect input validation).  
    \item \textbf{Impact:} the consequence of successful exploitation (e.g., privilege escalation, denial of service, data leakage).  
\end{itemize}}

\textcolor{black}{
These aspects represent the core information units that practitioners rely on to understand and mitigate vulnerabilities.}

\subsection{Pilot Experiments}~\label{sec_motivation_exp}
In addition to the takeaways from industry standards, we also conducted an empirical experiment to further explore the scope and depth of the aforementioned problems, revealing the extent to which reality falls short of industry standard expectations.

\textbf{Dataset.}
We first widely collect four popular web-based vulnerability repositories that are indexed by the CVE-ID: CVE (English), IBM (English), JVN (Japanese), and CNNVD (Chinese).
We then extract key aspects using the method described in Section~\ref{sec: extract}, which utilizes LLM with regularization templates derived from human guidelines. These templates, summarized from previous research~\cite{DBLP:journals/corr/abs-2002-08909,DBLP:conf/iclr/ZhouSHWS0SCBLC23, DBLP:journals/ieeesp/Massacci24}, replace the need for fixed examples in in-context learning, guiding the LLM to extract key aspects effectively. This approach helps us process TVDs from these four repositories spanning from 1999 to 2023.

\begin{figure}[tbp]
    \centering
    \includegraphics[width=0.85\textwidth]{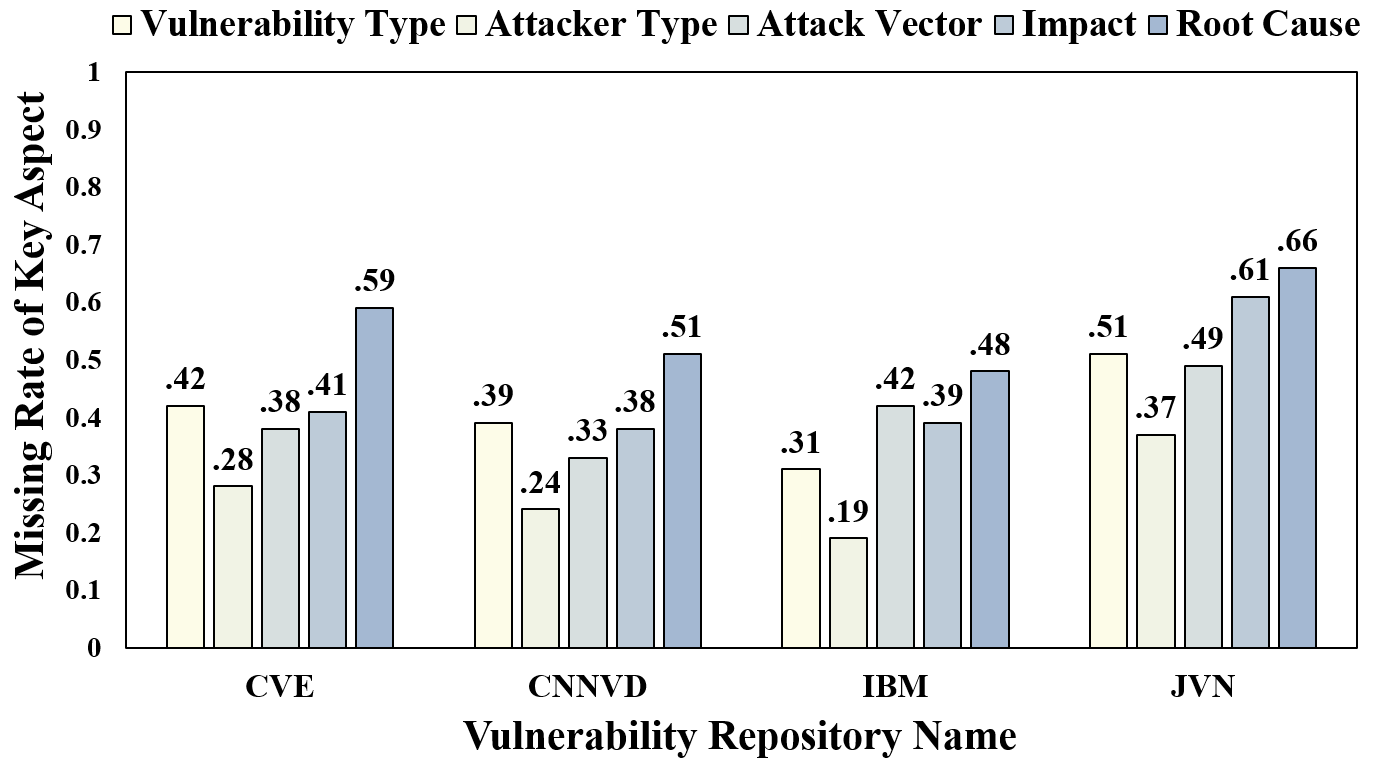}
    
    \caption{Missing rates on different vulnerability repositories.}
    \label{missingRate}
\end{figure}

\textbf{Information Missing. }
The missing rate for each key aspect is calculated by counting the occurrences and dividing by the total number of TVDs, then subtracting from 100\%. As shown in Figure~\ref{missingRate}, CVE and JVN have the highest missing rates, especially for \textit{Root Cause} (0.59 and 0.66). CNNVD shows the lowest missing rates, while IBM is moderate. These high missing rates, particularly for crucial aspects like \textit{Root Cause}, demonstrate a severe lack of essential information across repositories. This gap can hinder security engineers’ ability to fully understand vulnerabilities, resulting in delayed or ineffective responses.~\cite{DBLP:conf/icsm/SunXX0022, DBLP:conf/icse/SunX00023, DBLP:conf/compsac/GuoXCLBZ21} Standardized TVDs are needed to mitigate this issue.

\begin{table}[!t]
\centering
\caption{Missing and inconsistency rates for different key aspects.}
\label{inconsistencyRate}
\begin{tabular}{lccccc}
\toprule
   {\centering \tabincell{c}{Methods}} &{\centering \tabincell{c}{Vulnerability Type}} & {\centering \tabincell{c}{Attack Vector}} & {\centering \tabincell{c}{Attacker Type}} & Impact  &{\centering \tabincell{c}{Root Cause}} \\

\hline
\tabincell{c}{I. R.}    & \textbf{0.67} & \textbf{0.42} & \textbf{0.36}& \textbf{0.48} & 0.38  \\
\tabincell{c}{M. M.}    & 0.32 & 0.20 & 0.34& 0.38 & \textbf{0.49}  \\
\tabincell{c}{M. R.}   & 0.15 & 0.09 & 0.20& 0.17 & 0.23  \\

\bottomrule
\end{tabular}
\footnotetext{``I. R.'' refers to ``Inconsistency Rate'', ``M. M.'' represents ``Min Missing Rate in Single Repository'', and ``M. R.'' indicates ``Missing Rate''.}
% }
\end{table}

\textbf{Information Inconsistency.}
Previous studies have thoroughly examined the inconsistencies in key aspects like \textit{Affected Product, Component}, and \textit{Version}~\cite{DBLP:journals/compsec/SunOZLWZ23, DBLP:conf/ccs/QinXL23, DBLP:journals/tosem/SunXXLXZ24, DBLP:conf/uss/DongGCXZ019, DBLP:journals/tdsc/HeWZWZLY24}. These inconsistencies pose a significant challenge to security engineers as they lead to gaps in vulnerability understanding, miscommunication, and even incorrect threat assessments, potentially delaying critical remediation efforts. The severity of the issue is illustrated by the inconsistency rates ranging from 0.36 to 0.67 across different repositories (Table\ref{inconsistencyRate}), indicating that many key aspects, such as \textit{Vulnerability Type} and \textit{Attack Vector}, often differ between sources.

Moreover, merging data from multiple sources substantially reduces the missing rates for key aspects, enhancing the completeness of the information. For instance, before merging, the missing rate for \textit{Root Cause} in single repositories ranged from 0.49 (CNNVD) to 0.66 (JVN). After merging, the missing rate dropped to 0.23, a reduction of more than 50\%. Similarly, for \textit{Vulnerability Type}, the missing rate decreased from 0.32 to 0.15, and for \textit{Attack Vector}, from 0.20 to 0.09. This demonstrates that combining information from different repositories not only resolves a large portion of the missing information but also improves the overall reliability and accuracy of the TVDs. The substantial decrease in missing rates highlights the value of data consolidation in delivering more comprehensive and consistent vulnerability data.

Despite industry standards, current TVDs lack standardization, leading to high missing rates and inconsistencies across information sources. Guo et al.~\cite{DBLP:journals/tosem/GuoCXLBS22, DBLP:conf/compsac/GuoXCLBZ21, DBLP:conf/icsm/SunXX0022} highlights that such gaps can lead to increased error rates in vulnerability assessments, inefficiencies in remediation efforts, and potential misinterpretation of security risks. This increases labor costs, delays remediation, and raises the risk of misinterpretation. Therefore, we propose Digest Labels (DLs) that consolidates key aspects from multiple sources, facilitating practitioners in CVE-related downstream tasks.

%% file: 4_LabelDesign.tex
\begin{figure*}[htbp]
    \centering
    \includegraphics[width=1\textwidth]{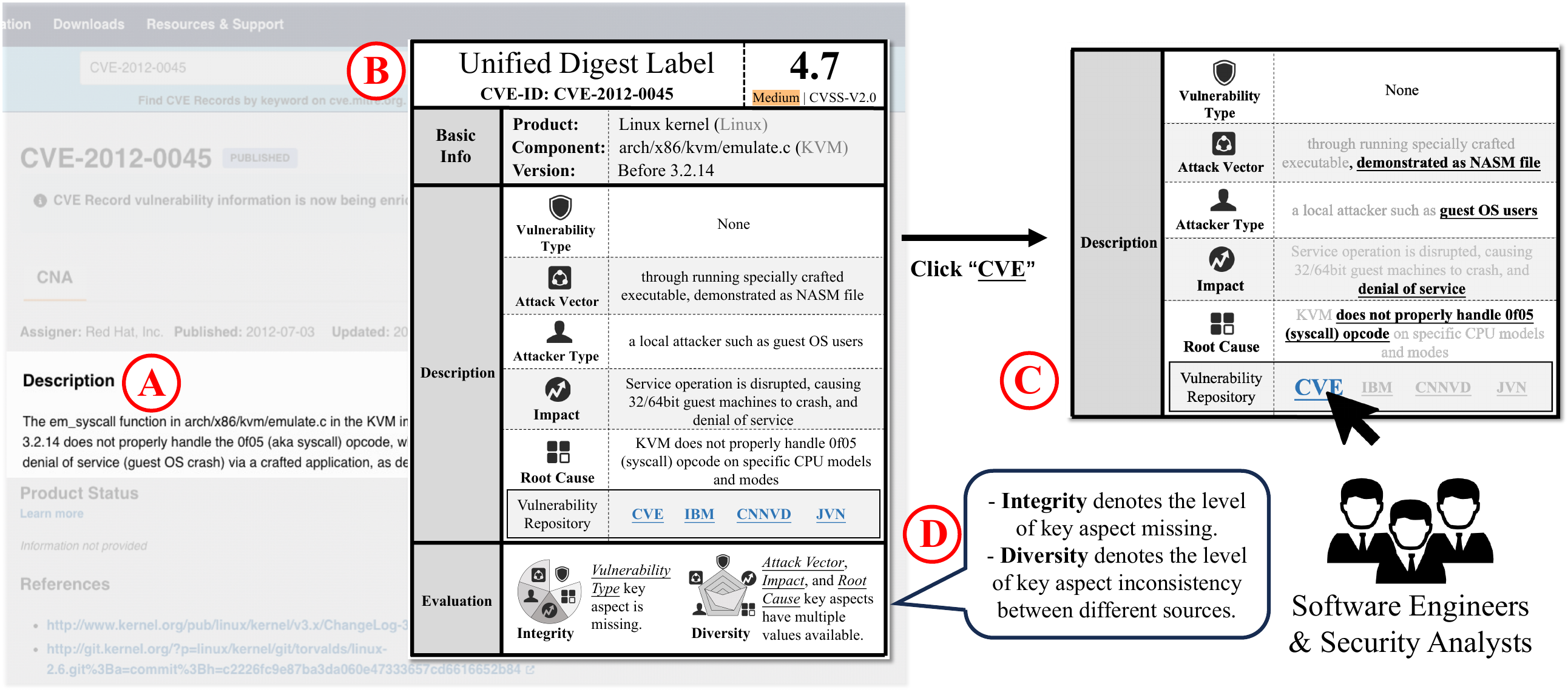}
    \caption{The application scenario of Digest Labels (DLs). A) is the textual vulnerability description (TVD) available in a CVE repository. B) is the design of proposed DLs. C) is the \textit{Description} section when the \textit{``CVE''} is selected. D) is the explanations of \textit{Evaluation} section.}
    \label{fig_UDL_design}
\end{figure*}  

\section{Design of TVD Digest Labels}
\label{labelDesign}

In the formative study, we discuss the long-standing challenges associated with CVE TVDs, such as the high rates of missing and inconsistent information. Current solutions often fall short in addressing these issues effectively. Our goal is to provide a more complete and standardized representation of vulnerability information for software engineers. Inspired by the concept of nutrition labels and their widespread application in various domains, we aim to adopt this approach to enhance the clarity and usability of TVD information.
While our formative study highlights the issues of missing and inconsistent information, our label design is motivated by the need to streamline the information reception process for security engineers. The \textit{Digest Labels} for CVE TVDs are designed to consolidate key aspects from multiple sources, standardize the presentation, and reduce the cognitive load on security engineers. This approach enables engineers to quickly and efficiently receive essential information, improving their understanding and response times.

\subsection{The Development of Nutrition Label in SE}
\subsubsection{Pre-millennium}
\textbf{Labels denotes Comments for \underline{Codes}.}
In the early stages of software engineering, the concept of labeling primarily manifest in code comments and documentation~\cite{boehm1988a, pressman2005software}. These labels provide information about inputs, outputs, code functionality, and dependencies, in a standardized manner, aiding developers to quickly understand and maintain the programs.

\subsubsection{Post-millennium}
\textbf{Labels denote Metadata for \underline{Software}.}
With the proliferation of open-source software and libraries, package managers such as Maven, NPM, and PyPI introduced metadata labels~\cite{microsoft2009maven, npm2014npm}. These labels include information on versioning, dependencies, and licenses, helping developers make more informed decisions when utilizing third-party libraries.
Recently, the Software Bills of Material (SBOM) attract considerable attention, considering as a crucial building block to ensure the transparency of software supply chains~\cite{xia2023empirical}.

\subsubsection{Present}
\textbf{Labels denote Information Transparency for \underline{Users and Developers}.}
Currently, labels serve to provide key information through standardization, helping users better understand and assess the dataset~\cite{pushkarna2022data}, model~\cite{mitchell2019model}, security~\cite{simko2019ask}, privacy~\cite{pan2023toward}, compliance~\cite{si2024solution}, and other important aspects of software systems. 
Standardized labels enhance transparency and accountability, improve user experience, and promote responsible and transparent development in the software engineering field.
As the modern software becomes larger and more sophisticated, developers share the same need as users to understand software information in a timely manner.

\subsection{Design of Digest Labels (DLs)}
\label{Visual Design}
Our DLs design aims to address the two major challenges identified in the formative study, missing and inconsistent key aspects in CVE TVD.
The designed label is shown in (a) in Figure~\ref{fig_UDL_design}.

The DLs is indexed by the CVE-ID, as a unique identifier to indicate CVEs.
Common Vulnerability Scoring System (CVSS) is a method used to supply a qualitative measure of severity~\cite{nist}, and software engineers highly depend on the scores to estimate the severity of the problem and decide the priority.
The DLs contains three sections, Basic Information, Description, and Evaluation.
Basic Information section presents the \textit{Product, Component}, and \textit{Version}. 
For any inconsistencies in the software name and version, the most frequent aspects are displayed in solid black, while the remaining inconsistent aspects are shown in lighter grey as reference. 
The second section is the key aspects, where the default view shows the merged results using the method from Section~\ref{approa: merge}. 
The third section is the evaluation of the TVDs. \textit{Integrity} denotes the level of key aspect missing and \textit{Diversity} denotes the level of key aspect inconsistency between different sources.

To better demonstrate the inconsistencies in information across different repositories, we design a interactive presentation for Description section. 
This design allows viewers to click different information sources in the ``Vulnerability Repository''. 
For example, clicking the ``CVE'' option will show the key aspects that mainly extracted and contributed by the ``CVE'', as shown in Figure~\ref{fig_UDL_design}(c).

%% file: 5_Approach_V1.tex
\section{Domain-Constrained Synthesis}
\label{approach}
\begin{figure*}[]
    \centering
    \includegraphics[width=0.99\textwidth]{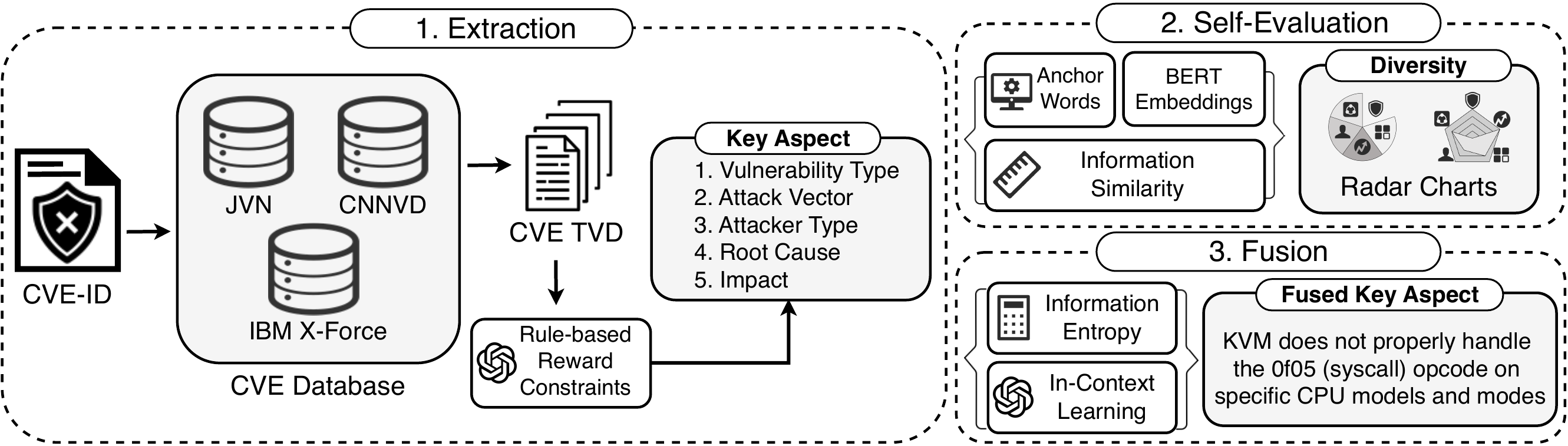}
    \caption{The overview of framework.}
    \label{overview}
\end{figure*}

The domain-constrained synthesis framework for key aspects is illustrated in Figure~\ref{overview}. It consists of three main components: key aspect extraction, self-evaluation and key aspect fusion.

\subsection{Extraction through Rule-Based Reward Constraints}
\label{sec: extract}

Extracting key aspects from TVDs is challenging due to the varied expressions in which information is presented. Traditional methods~\cite{DBLP:journals/corr/abs-2405-07430, DBLP:journals/tosem/GuoCXLBS22, DBLP:conf/kbse/YitagesuXZ00H21} based on In-Context Learning focus on identifying structured terms but struggle with the diversity of key aspect expressions in vulnerabilities~\cite{DBLP:journals/corr/abs-2002-08909, DBLP:conf/iclr/ZhouSHWS0SCBLC23, DBLP:journals/ieeesp/Massacci24}, leading to missed critical details.

Rule-based Reward is proposed by~\cite{rule_based_rewards_2023}, to address data annotation issues by synthesizing training data with AI feedback, combined with human data, for supervised fine-tuning (SFT) and reward model (RM) training. 
The essence of Rule Based Reward lies in using general human guidelines to direct AI generation. We apply the Rule Based Reward to prompts by using general human guidelines as substitutes for examples to guide the LLM. 
In extracting key aspects of vulnerabilities, these general human guidelines are regularization templates, which are summarized by previous researches~\cite{DBLP:journals/tosem/GuoCXLBS22}. 
These templates, acting as domain-specific constraints, explicitly define the essential elements of vulnerabilities (e.g., ``opcode'', ``specific CPU models'').
By replacing in-context learning examples with LLM extraction constitution, we can overcome the limitation of example numbers. This approach uses human experience to guide the extraction process of LLM. 
The list of regularization templates for key aspects is available in our artefact repository.

\subsection{Self-Evaluation Though Anchor Words Constraints}
\label{app: tvd eval}

The Evaluation section in DLs aims to provide a comprehensive quality assessment of the TVDs. 
While prior studies~\cite{DBLP:journals/compsec/SunOZLWZ23, DBLP:conf/ccs/QinXL23, DBLP:journals/tosem/SunXXLXZ24, DBLP:conf/uss/DongGCXZ019, DBLP:journals/tdsc/HeWZWZLY24} focus on statistical analysis, we evaluate key aspects based on two criteria: information integrity and diversity (Figure~\ref{fig_UDL_design}).

\textbf{Integrity. } 
Information integrity assesses whether all key aspects are present. Complete key aspects enhance user understanding of vulnerabilities. If all aspects are present, we display a pie chart illustrating the five key aspects and a message stating ``There are no missing key aspects''. 
On the contrary, missing aspects result in blank segments on the pie chart, and curated by side.
The calculation of Integrity is merely the difference in the number of key aspects. Therefore, the calculation process relies on the extraction of key aspects, and the accuracy of the Integrity depends on the accuracy of key aspect extraction.

\begin{algorithm}

\caption{Distance Calculation of Key Aspects}
\begin{algorithmic}[1]
\Require \text{Aspects} = $[a_1, a_2, \ldots, a_n]$
\Function{LLM\_extract}{a}
    \State \text{prompt} = ``Extract computer specific terms from sentence: ''
    \State \text{prompt} = \text{prompt} + a
    \State \Return \text{LLM}(\text{prompt})
\EndFunction
\State \text{similar} = $[]$
\For{each $i$ in \text{Aspects}}
    \For{each $j$ in \text{Aspects}}
        \State \text{temp1} = \text{BertVec}(\Call{LLM\_extract}{i})
        \State \text{temp2} = \text{BertVec}(\Call{LLM\_extract}{j})
        \State \text{sim} = \text{Cosine}(\text{temp1}, \text{temp2})
        \State \text{similar.append}(\text{sim})
    \EndFor
\EndFor
\State \text{res} = \text{average}(\text{similar})
\State \Return \text{res}
\end{algorithmic}
\label{alg:sim}
\end{algorithm}

\textbf{Diversity.} 
Information diversity measures overlap among key aspects.
High diversity, indicated by low overlap, offers valuable unique information.
We utilize BERT sentence embeddings to represent the semantic vectors corresponding to each key aspect, using SpaCy's\footnote{\url{https://spacy.io/}} Transformer module (spacy-transformers) to generate BERT embeddings for output.
We calculate cosine similarity~\cite{DBLP:conf/icann/LuoZXWRY18} between vectors.
However, inconsistencies can lead to significant distances between vectors for synonymous key aspects.
While expressions may differ, computer-specific terms—acting as semantic anchors—remain constant.
Although there are dictionaries for general computer terminology\footnote{\url{https://github.com/sohale/cs-glossaries}}, there is no dedicated dictionary for the vulnerability domain. The uniqueness of computer-specific terms makes it challenging to apply general-purpose computer dictionaries in this context. For example, terms like “buffer overflow” and “SQL injection” have specific meanings in vulnerability contexts that may not be adequately captured in broader computer science glossaries.
To impose constraints in the similarity computation, we employ LLM to extract domain-specific anchor words from each key aspect. These anchor words serve as constraint points that guide the semantic similarity comparison.
The process is detailed in Algorithm~\ref{alg:sim}, with $Aspects$ representing all key aspects under consideration.

The information dispersion value ranges from $[0,1]$, with a $0.2$ mapping to the 5-point Likert.
Both information dispersion and information completeness are represented using radar charts, located in the upper right corner of the label.
After calculating the distances between the key aspects, we compute the distances for different values within the same key aspect type and then find the average distance. Once the distances for all five key aspect values are calculated, we plot these values on a radar chart. Any key aspect types with values greater than 2 are also noted in text.

\subsection{Fusion though Entropy Constraints}
\label{approa: merge}

\begin{figure}[t]
    \centering
    \includegraphics[width=0.85\textwidth]{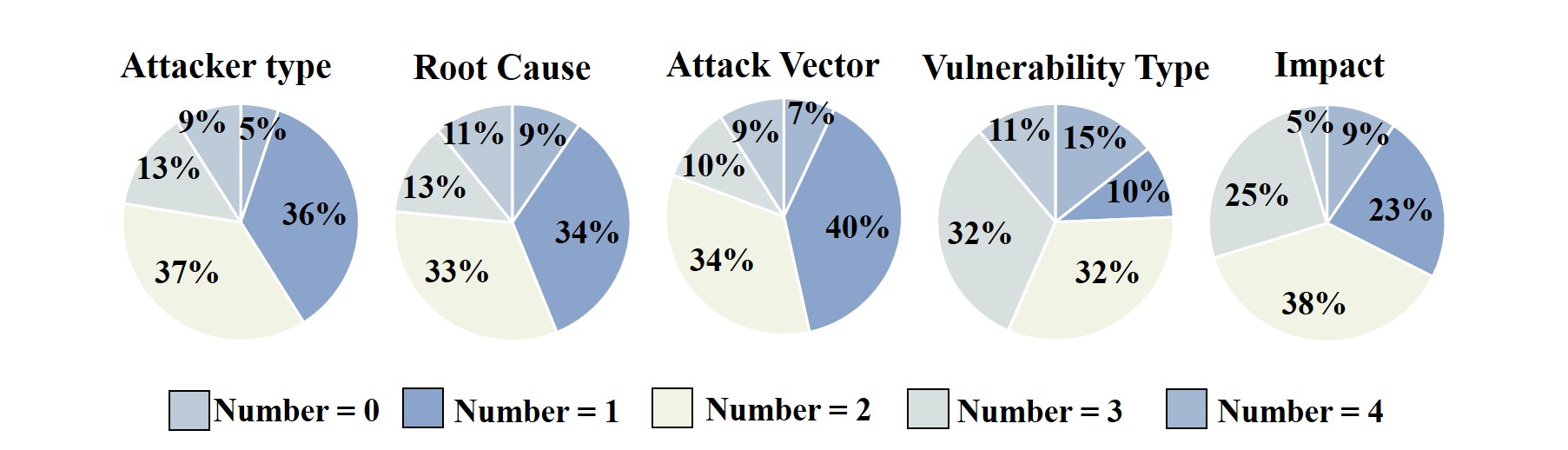}
    \caption{Numbers proportion of key aspect values}
    \label{count_number}
\end{figure} 

As shown in Table~\ref{inconsistencyRate}, the inconsistency rate for key aspects in TVDs ranges from 0.36 to 0.67, indicating that nearly half have more than one value. 
We further statistically count the number of values for each key aspect across four vulnerability repositories (CVE, IBM, CNNVD, and JVN), finding that most of them have 2 to 3 different values, as illustrated in Figure~\ref{count_number}. 
Additionally, we conduct a further word count analysis, revealing that the average lengths for key aspects: \textit{Vulnerability Type} (5.1 words), \textit{Attacker Type} (3.8 words), \textit{Attack Vector} (13.6 words), \textit{Impact} (9.7 words), and \textit{Root Cause} (8.4 words). 
Thus, security engineers read an average of 24 words per key aspect type, totaling around 120 words for all descriptive aspects, many of which are similar, plus the additional time to cross-reference. 
Our fusion approach seeks to maximize \textbf{non-consensus information} from various sources, refining the presentation unlike the structured information in TVDs.
Information Entropy~\cite{shannon1948mathematical} is an indicator to measure information richness in a sentence without being affected by expression, and we utilize it to achieve the maximal non-consensus information.
By maximizing entropy with an effective aggregation method, we ensure comprehensive retention of information, aiming to avoid critical details being lost or duplicated. 
By preserving their diversity and clarity, our approach could be useful for vulnerability analysis, management, and CVE-related downstream tasks. 

Specifically, the key aspect information entropy calculation method is shown in Equation~\ref{entropy_formula}.
Let $s1,s2..sn$ denote $n$ key aspects that need to be merged. The function $\text{connect}(s_1,s_2,...,s_n)$ represents the concatenation of $s1,s2..sn$. $\text{count}(w_i)$ indicates the number of times the word $w_i$ appears in set $S$.
\begin{equation}
S = \text{connect}(s_1,s_2,...,s_n)
\end{equation}
\begin{equation}
\label{entropy_formula}
E = \sum_{i=1}^{len(S)} \left( \frac{\text{count}(w_i)}{len(S)}\log \frac{\text{count}(w_i)}{len(S)} \right)
\end{equation}
%说LLM遇到的问题
After calculating the information entropy, we need to combine TVDs into semantically readable and syntactically correct descriptions to display.
LLM can leverage their sentence rewriting abilities to combine sentences~\cite{DBLP:conf/aaai/0004LHZLTCM24, DBLP:journals/corr/abs-2401-11911}. 
However, when merging content between sentences, LLM tend to merge all sentences, even if there is no semantic overlap. 
This increase in sentence complexity can hinder the readability of security engineers.
%我们的思路
To mitigate this, we introduce entropy-based constraints into the prompt design. By heuristically embedding the calculated information entropy, we guide the LLM to focus on the most informative and non-redundant elements during fusion~\cite{DBLP:journals/corr/YuCLL16, DBLP:journals/sigir/ZhaiL17, DBLP:journals/entropy/BentzACF17, DBLP:journals/entropy/Levshina22}.
Even when different expressions describe the same key aspect, the entropy constraint preserves the most relevant content, enforcing a form of information-aware merging.
Specifically, we use word-frequency-based information entropy to quantify the richness of each candidate and apply this value as a constraint that discourages over-merging and encourages information preservation.
We compute entropy among key aspects and provide these values to the LLM to assist in sentence merging, as shown in Algorithm~\ref{mergePrompt}.
The $sentence_list$ consists of examples randomly selected by humans, and the $merge_result$ is manually constructed for supervision.

\begin{algorithm}

\caption{Key Aspect Merging by Information Entropy}

\label{mergePrompt}
\begin{algorithmic}[1]
\State \textbf{Input:} sentence\_list = $[s_1, s_2, \ldots, s_n]$
\State \textbf{Output:} merge\_result

\State \textbf{Example1:}
\State \hspace{1em} \textbf{Task:} "Based on information entropy, I can merge sentence\_list."
\State \hspace{1em} \textbf{Sentence\_list:} $[k_1, k_2, \ldots, k_n]$
\State \hspace{1em} \textbf{Information entropy:} $E$ of sentence\_list
\State \hspace{1em} \textbf{Merge result:} (merge of sentence\_list)
\State \textbf{Example2:} ...
\State \textbf{Example3:} ...
\State \textbf{Task Description:}
\State \hspace{1em} \textbf{Task:} "Based on information entropy, I can merge sentence\_list."
\State \hspace{1em} \textbf{Sentence\_list:} $[s_1, s_2, \ldots, s_n]$
\State \hspace{1em} \textbf{Information entropy:} $E$ of sentence\_list
\State \hspace{1em} \textbf{Merge result}
\end{algorithmic}
\end{algorithm}

Our core objective is to ensure users receive the maximum amount of valuable information without redundancy. We achieve this through two key strategies:
\begin{itemize}
    \item a) Information presentation. As shown in Figure~\ref{fig_UDL_design}, part B offers four different repository options. Users can click on each repository to view the specific information it provides. To avoid wasting users' time, redundancy must be minimized through content constraints.
    \item b) Application of Information Entropy as Constraints. During the information fusion process, entropy values act as soft constraints to regulate LLM behavior, encouraging minimal redundancy and maximizing informativeness. Unlike the diversity metric in Section~\ref{app: tvd eval}, which measures semantic difference, entropy captures content richness while ignoring superficial syntactic variations. This constraint mechanism ensures that only meaningful differences are preserved in the final output.
\end{itemize}

\textcolor{black}{
\subsection{Synthesis Baseline: Vanilla LLM Workflow}
\label{sec:baseline}
To provide a fair comparison, we introduce a baseline workflow that simulates how practitioners might directly use LLM without additional domain-specific strategies. 
As shown in Algorithm~\ref{res: prompt}, the vanilla workflow only involves prompting the LLM to (a) summarize TVDs for a given CVE-ID, and (b) directly extract, merge, and evaluate key aspects from the summarized text. 
This baseline reflects the most straightforward usage of LLM, where no reward-based constraints, anchor words, or entropy-based fusion strategies are applied. 
We use this baseline to evaluate whether our proposed framework provides substantial improvements over the naive application of LLM.}

\begin{algorithm}

\caption{Label Generation Prompt}
\label{res: prompt}
\begin{algorithmic}[1]
\State \textbf{Input:} Three randomly selected CVE-IDs
\State \textbf{Output:} Extracted Key Aspects, Merged Key Aspects, TVD Evaluation
\State \textbf{Task Description:}
\State \hspace{1em} \textbf{Task:} ``Based on the provided examples, please return:''
\State \hspace{2em} \textbf{CVE-ID:} (CVE-ID)
\State \hspace{2em} \textbf{TVDs:} \{(TVD from CVE), (TVD from CNNVD),...\}
\State \hspace{2em} \textbf{Please Return:} Extracted Key Aspects, Merged Key Aspects, TVD Evaluation.
\end{algorithmic}
\end{algorithm}

\textcolor{black}{    
To further strengthen the baseline, we also consider a \textit{Chain-of-Thought (CoT) prompting} variant, which encourages the LLM to reason step by step before generating the final outputs. 
This variant builds on the same input/output format of Algorithm~\ref{res: prompt}, but modifies the task instruction to explicitly require intermediate reasoning steps, as shown inAlgorithm~\ref{alg:cot}
}

\begin{algorithm}
\caption{Chain-of-Thought Prompting Baseline}
\label{alg:cot}
\begin{algorithmic}[1]
\State \textbf{Input/Output:} Same as Algorithm~\ref{res: prompt}.
\State \textbf{Task Description:}
\State \hspace{1em} \textbf{Instruction:} ``Think step by step before answering.''
\State \hspace{2em} Step 1: Read all provided TVDs carefully.
\State \hspace{2em} Step 2: Identify candidate key aspects from each TVD.
\State \hspace{2em} Step 3: Compare candidate aspects and remove duplicates or conflicts.
\State \hspace{2em} Step 4: Return final merged aspects and evaluation.
\State \hspace{2em} \textbf{Please Return:} Same as Algorithm~\ref{res: prompt}.
\end{algorithmic}
\end{algorithm}

%% file: 7_Results.tex
\section{Evaluation} 
\label{result}
This section evaluates the performance of the generation framework and its usefulness.

\subsection{Dataset} 
\label{dataset}
We collected 289,105 CVE-IDs from 1999 to 2023, resulting in 956,619 TVDs from four repositories: CVE, IBM X-Force (IBM), CNNVD, and JVN. A representative subset of 384 \textcolor{black}{CVE-IDs} was randomly sampled with a 95\% confidence level and 5\% margin of error.

\textcolor{black}{
To further demonstrate the representativeness of our sample, we provide its distribution across time and software types. As shown in Table~\ref{tab:sample_dist}, the 384 CVE-IDs are evenly distributed over the years from 1999 to 2023, and they cover a diverse set of software domains including operating systems, web applications, databases, libraries, and middleware. This distribution indicates that our sample captures both temporal and domain diversity, thereby supporting the generalizability of our evaluation.
}

\begin{table}[t]
\centering
\caption{Distribution of the sampled 384 CVE-IDs across years and software types.}
\label{tab:sample_dist}
\begin{tabular}{lccc}
\toprule
\textbf{Category} & \textbf{Subcategory} & \textbf{\# Samples} & \textbf{Proportion} \\
\midrule
\multirow{5}{*}{Year Range} 
  & 1999--2005   & 42  & 10.9\% \\
  & 2006--2010   & 56  & 14.6\% \\
  & 2011--2015   & 73  & 19.0\% \\
  & 2016--2020   & 102 & 26.6\% \\
  & 2021--2023   & 111 & 28.9\% \\
\midrule
\multirow{6}{*}{Software Type} 
  & Operating Systems       & 88  & 22.9\% \\
  & Web Applications/Frameworks & 105 & 27.3\% \\
  & Database Systems        & 41  & 10.7\% \\
  & Programming Languages/Libraries & 52 & 13.5\% \\
  & Middleware/Virtualization & 46 & 12.0\% \\
  & Others (IoT/Drivers)    & 52  & 13.5\% \\
\midrule
\multicolumn{2}{l}{\textbf{Total}} & \textbf{384} & \textbf{100\%} \\
\bottomrule
\end{tabular}
\end{table}

\textcolor{black}{
\subsection{Construction of Ground Truth}
\label{ground_truth}
}

\textcolor{black}{
As there is no existing ground truth for our novel task of TVD synthesis, we manually annotated the dataset to create the benchmark for evaluation. The annotation process was conducted as follows:
}

\textcolor{black}{
\textbf{Annotator Recruitment and Qualifications:}
We recruited five evaluators, all of whom are professionals or researchers specializing in software vulnerabilities. Each evaluator had either participated in software vulnerability projects for more than three years or had at least two years of experience and had published papers on security vulnerabilities.
}

\textcolor{black}{
\textbf{Annotation Task:}
For each of the 384 CVE-IDs in our sample, evaluators were asked to label the output of each sub-task in our framework as \textit{True} or \textit{False} based on the following criteria:
\begin{itemize}
    \item For Key Aspect Extraction (Section~\ref{Extracting of Key Aspect}), \textit{True} means that the extracted key aspect is correct and appears in the source TVD, while \textit{False} indicates an incorrect or missing extraction.
    \item For Key Aspect Fusion (Section~\ref{res: merge}), \textit{True} signifies that the merged key aspects have neither lost essential information nor added extraneous information, whereas \textit{False} indicates a loss or unjustified addition of information.
    \item For TVD Evaluation (Section~\ref{res: Score}), \textit{True} indicates that the algorithm's computed dispersion level (among the 5 levels) matches the human judgment, while \textit{False} indicates a discrepancy.
\end{itemize}}

\textcolor{black}{
\textbf{Calibration and Adjudication Process:}
Before the formal annotation, we conducted a calibration stage in which the five evaluators independently annotated a small subset of CVEs. During this stage, Fleiss' Kappa values in some cases fell below 0.2, which is generally regarded as poor agreement beyond chance~\cite{mchugh2012interrater}, reflecting the difficulty and subjectivity of certain samples. To resolve these disagreements, the annotators engaged in structured discussions until they reached a shared understanding of the labeling criteria. The annotation guidelines were refined accordingly to reduce ambiguity. After calibration, the evaluators proceeded to annotate the full dataset, and any residual disagreements were resolved by majority vote, a standard practice in annotation studies~\cite{bornmann2011manuscript, hallgren2012computing}.
}

\textcolor{black}{
\textbf{Final Inter-Annotator Agreement:}
This process resulted in a high-quality ground truth dataset with substantial agreement. The final Fleiss' Kappa values for the tasks were as follows:
\begin{itemize}
    \item \textbf{0.806} for Key Aspect Extraction (Figure~\ref{fig: key aspect extraction}),
    \item \textbf{0.781} for TVD Evaluation (Figure~\ref{fig: key aspect eval}),
    \item \textbf{0.835} for Key Aspect Fusion (Table~\ref{res: key aspect merge}),
    \item \textbf{0.784} for the overall framework evaluation (Table~\ref{res: overview}).
\end{itemize}
These values indicate a substantial level of agreement, ensuring the reliability of our ground truth for the subsequent evaluation of the framework's performance.
}

\textcolor{black}{
\subsection{Experimental Setup}
\label{sec:exp_setup}
To ensure the reproducibility of our experiments, we provide a detailed description of our LLM configurations and prompting strategies. All experiments were conducted using the ERNIE-3.5 API\footnote{\url{https://yiyan.baidu.com/}} as the primary large language model, with a fixed parameter configuration: model version \texttt{ernie-3.5-8k}, temperature=0.8, top-p=1.0, frequency penalty=0, presence penalty=0, and max output tokens=4,096. We follow the common practices of LLM tasks for those settings. These hyperparameters were kept constant across all tasks.
}

\textcolor{black}{
Our prompting strategy integrates domain-specific constraints into each sub-task. For Key Aspect Extraction (Section~\ref{sec: extract}), we employ \textit{rule-based rewards} by embedding human-summarized regularization templates~\cite{DBLP:journals/tosem/GuoCXLBS22} as guidelines within the prompt, instructing the model to adhere to these domain-specific structures for identification. For Self-Evaluation (Section~\ref{app: tvd eval}), the prompt is designed to extract semantic anchor words using the direct instruction: ``Extract computer-specific terms from the following sentence: [sentence]''. For Key Aspect Fusion (Section~\ref{approa: merge}), we use a few-shot learning prompt (Algorithm~\ref{mergePrompt}) where the calculated information entropy of the input sentences is provided as a key constraint to guide the model towards merging for maximal non-redundant information.
}

\subsection{Performance of Synthesis}
We first individually evaluate three sub-tasks for each section, and then assess the overall performance of the framework in solving the two challenges: information missing and inconsistency.
Notably, we select Ernie-3.5\footnote{\url{http://research.baidu.com/Blog/index-view?id=185}} as our default LLM choice due to its performance, accessibility, and cost-effectiveness.

\subsubsection{Key Aspect Extraction}
\label{Extracting of Key Aspect}

\begin{figure}[t]
\label{fig: key aspect extraction}
    \centering
    \includegraphics[width=0.85\textwidth]{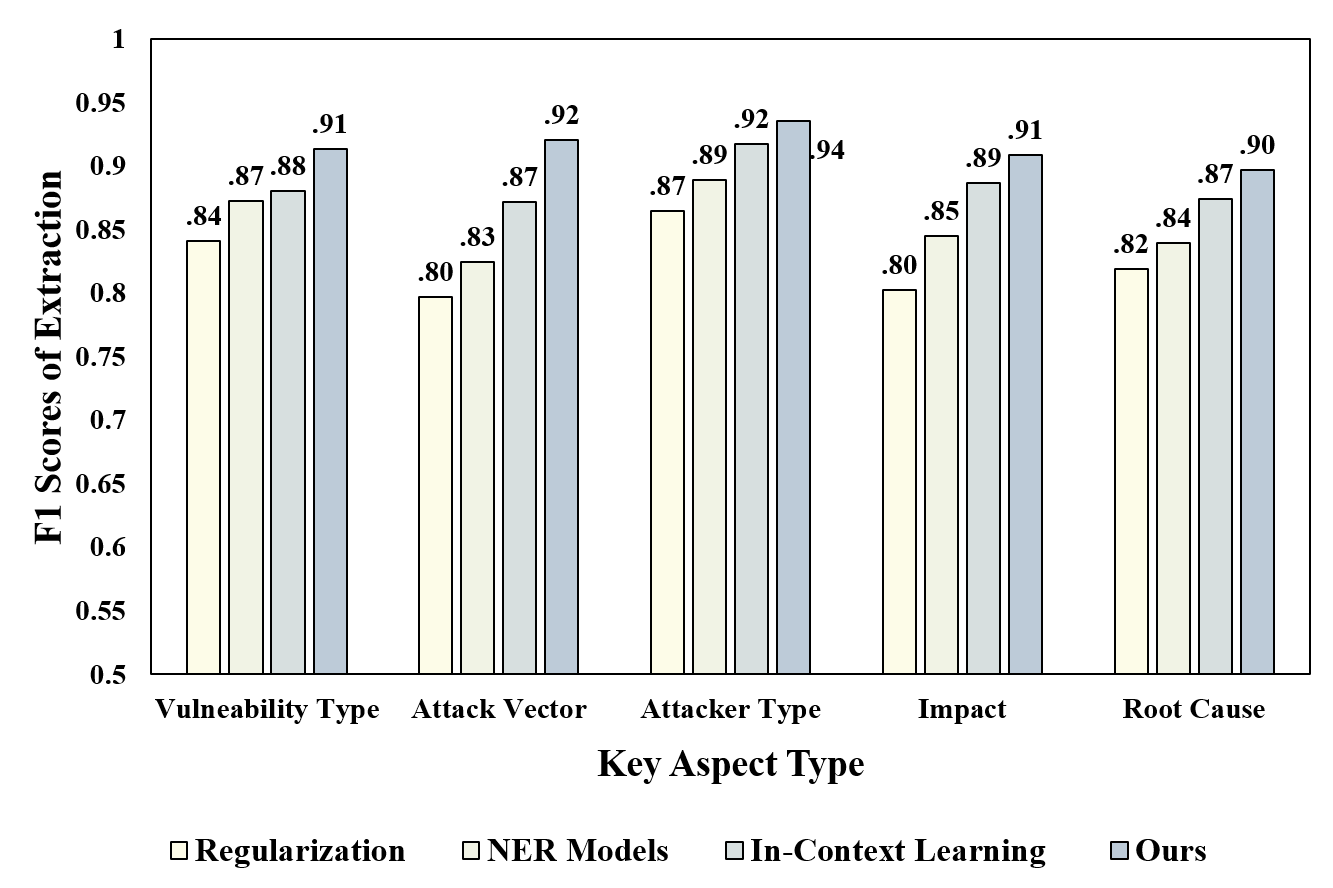}
    
    \caption{F1 Score of key aspects extraction by different methods.}

\end{figure}

Figure~\ref{fig: key aspect extraction} compares the performance of different key aspect extraction methods, and we select the following baseline methods:
\begin{itemize}
\item \textbf{[Regularization]}:  Guo et al.~\cite{DBLP:journals/tosem/GuoCXLBS22} summarize TVDs regularization templates and extract key aspects based on these templates.
\item \textbf{[NER Models]}: 
Named-entity recognition (NER) models for vulnerability~\cite{DBLP:journals/tosem/YitagesuXZFLH23} use unsupervised clustering for the automated classification of key aspects. 
\item \textbf{[In-context Learning]} Example-based in-context learning~\cite{DBLP:journals/corr/abs-2002-08909,DBLP:conf/iclr/ZhouSHWS0SCBLC23, DBLP:journals/ieeesp/Massacci24} provide several examples in the prompt and then query the LLM to obtain results.
\end{itemize}

Our rule-based reward method achieves the highest F1 scores across all five key aspects: 0.91 for \textit{Vulnerability Type}, 0.92 for \textit{Attack Vector}, 0.94 for \textit{Attacker Type}, 0.91 for \textit{Impact}, and 0.90 for \textit{Root Cause}. In comparison, the in-context learning method performs slightly worse, followed by NER models and the regularization method. Overall, our approach demonstrates superior performance in extracting key aspects.

\subsubsection{Key Aspect Self-Evaluation}
\label{res: Score}
We choose two widely-adopted baseline methods to compare with our approach: a) Using BERT similarity directly as the distance between key aspects without employing a multi-agent approach; b) Glossary lists about software vulnerability. 
Specifically, we obtain the glossary lists by searching keywords ``computer science glossary'' (dubbed as CSG glossary) and ``IT terminology list'' (dubbed as ITL glossary), to collect relevant resources and combine them as glossary lists. Both lists are available at our artefact repository.
If words or phrases from the TVD appear in these glossaries, we then extract and replace them, i.e., substituting lines 2-4 of Algorithm~\ref{alg:sim} with the glossary extraction method.

\begin{figure}[tbp]
\label{fig: key aspect eval}
    \centering
    \includegraphics[width=0.85\textwidth]{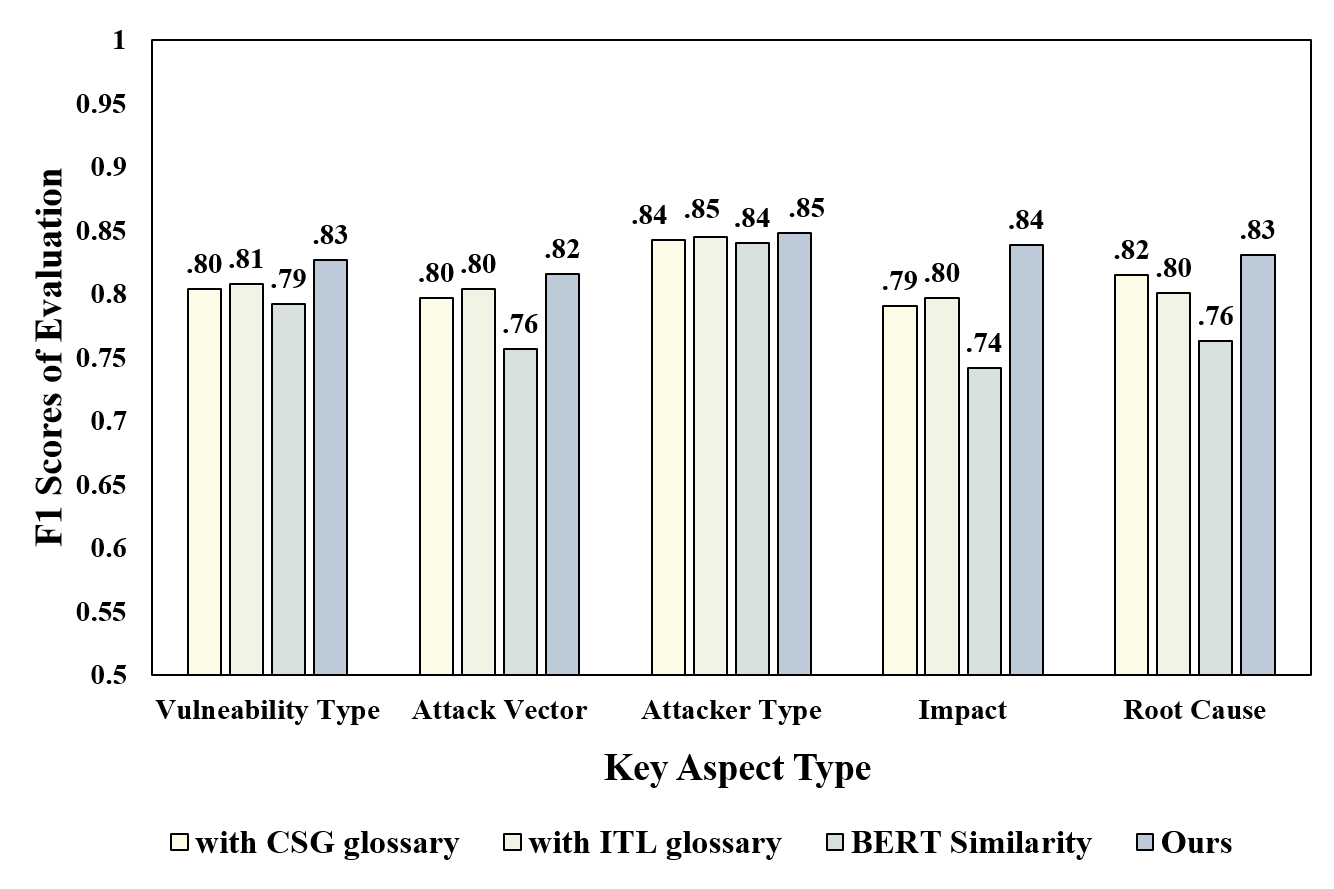}
    \caption{F1 Score of key aspects self-evaluation by different methods.}

\end{figure} 
Figure~\ref{fig: key aspect eval} compares the evaluation F1 score of TVD using different methods across five key aspects. Our method achieves the highest F1 score in all categories, with scores ranging from 0.82 to 0.85. The methods using CSG and ITL glossaries perform slightly lower, with accuracies between 0.79 and 0.84. The BERT similarity method shows the lowest F1 score, particularly in the \textit{Attack Vector} and \textit{Impact}. This highlights the effectiveness of our approach.

\subsubsection{Key Aspect Fusion}
\label{res: merge}

Considering the powerful sentence processing capabilities of LLM, we use an LLM-based in-context learning for key aspect fusion. 
We compare our method with two benchmarks: 1) do not provide constraints in prompts (i.e., removing the ``Information entropy'' from Algorithm~\ref{mergePrompt}), and 2) using BERT semantic similarity between key aspects as information content prompt (i.e., replacing the ``Information entropy'' in Algorithm~\ref{mergePrompt} with BERTScore similarity).

\begin{table}[t]
\centering

\caption{The F1 Score of key aspects fusion.}
\label{res: key aspect merge}
\begin{tabular}{lccccc}
\toprule
   {\centering \tabincell{c}{Methods}} &{\centering \tabincell{c}{Vulnerability Type}} & {\centering \tabincell{c}{Attack Vector}} & {\centering \tabincell{c}{Attacker Type}} & Impact  &{\centering \tabincell{c}{Root Cause}} \\
\midrule
\tabincell{c}{No Heuri.}    & 0.84 & 0.81 & 0.91& 0.83 & 0.83  \\
\tabincell{c}{BERT Sim.}   & 0.84 & 0.79 & 0.90& 0.84 & 0.82  \\
\tabincell{c}{Ours}    & \textbf{0.90} & \textbf{0.83} & \textbf{0.92}& \textbf{0.85} & \textbf{0.85}  \\
\bottomrule
\end{tabular}

\end{table}

Figure~\ref{res: key aspect merge} shows that our method, with information entropy as constraints, achieves high F1 scores across all five key aspects: 0.90 for \textit{Vulnerability Type}, 0.83 for \textit{Attack Vector}, 0.92 for \textit{Attacker Type}, 0.85 for \textit{Impact}, and 0.85 for \textit{Root Cause}. In comparison, the method without an information content prompt and the BERT Similarity method show lower F1 Scores, demonstrating that incorporating information entropy significantly improves key aspect fusion.

\textcolor{black}{
\subsubsection{Overall Performance}
\label{sec:overall_eval}
Having defined the baseline workflow in Section~\ref{sec:baseline}, we now compare its performance with our proposed framework. 
Table~\ref{res: overview} presents the results across the three sub-tasks: extraction, fusion, and self-evaluation. 
In addition to the vanilla LLM workflow, we also tested a stronger prompting variant using chain-of-thought (CoT) prompting as an alternative baseline. 
While CoT prompting improves certain sub-tasks slightly (e.g., fusion), it still shows limitations, particularly in overall consistency and self-evaluation (average F1=0.84). 
In contrast, our framework consistently outperforms both baseline variants, with ERNIE achieving the best overall performance (average F1=0.87). 
These results confirm that the domain-constrained design substantially improves the ability of LLM to handle information missing and inconsistency.
}

\begin{table}[]
  \centering
\caption{The overall performance on tackling the information missing and information inconsistency in CVE TVDs.}    
  \label{res: overview}
  \begin{tabular}{c|lcccc}
  \toprule
   & Methods & Extract & Fusion & Self-evaluation & Average \\
  \midrule
  \multirow{4}{*}{F1} 
    & ERNIE + Alg.~\ref{res: prompt} & 0.88 & 0.84 & 0.76 & 0.82 \\
    & ERNIE + Alg.~\ref{alg:cot}     & 0.87 & 0.86 & 0.78 & 0.84 \\
    & Ours (GPT-3.5)  & 0.90 & 0.88 & 0.82 & 0.86 \\
    & Ours (ERNIE)    & \textbf{0.92} & \textbf{0.87} & \textbf{0.83} & \textbf{0.87} \\
  \midrule
  \multirow{4}{*}{Accuracy} 
    & ERNIE + Alg.~\ref{res: prompt} & 0.80 & 0.75 & 0.70 & 0.75 \\
    & ERNIE + Alg.~\ref{alg:cot}     & 0.79 & 0.77 & 0.72 & 0.76 \\
    & Ours (GPT-3.5)  & 0.83 & 0.78 & 0.74 & 0.78 \\
    & Ours (ERNIE)    & \textbf{0.86} & \textbf{0.81} & \textbf{0.80} & \textbf{0.82} \\
  \midrule
  \multirow{4}{*}{Precision} 
    & ERNIE + Alg.~\ref{res: prompt} & 0.85 & 0.78 & 0.72 & 0.78 \\
    & ERNIE + Alg.~\ref{alg:cot}     & 0.84 & 0.80 & 0.74 & 0.79 \\
    & Ours (GPT-3.5)  & 0.88 & 0.83 & 0.75 & 0.82 \\
    & Ours (ERNIE)    & \textbf{0.90} & \textbf{0.81} & \textbf{0.79} & \textbf{0.83} \\
  \midrule
  \multirow{4}{*}{Recall} 
    & ERNIE + Alg.~\ref{res: prompt} & 0.82 & 0.80 & 0.74 & 0.79 \\
    & ERNIE + Alg.~\ref{alg:cot}     & 0.81 & 0.82 & 0.76 & 0.80 \\
    & Ours (GPT-3.5)  & 0.85 & 0.80 & 0.78 & 0.81 \\
    & Ours (ERNIE)    & \textbf{0.89} & \textbf{0.84} & \textbf{0.81} & \textbf{0.85} \\
  \bottomrule
  \end{tabular}
\end{table}

\begin{table*}[]
\centering
\small
\caption{Performance across different key aspects with different LLM}
\label{tab:key_aspect_llm_full}
\begin{tabular}{lll|ccccc}
\toprule
\textbf{LLM} & \textbf{Method} & \textbf{Metric} 
& \textit{Att. Vec.} & \textit{Ro. Cau.} & \textit{Imp.} & \textit{Vul. Typ.} & \textit{Att. Typ.} \\
\midrule
\multirow{8}{*}{GPT-3.5} 
  & \multirow{4}{*}{Alg.~\ref{res: prompt}} 
    & Acc  & 0.78 & 0.71 & 0.75 & 0.82 & 0.87 \\
  & & Prec & 0.80 & 0.66 & 0.70 & 0.79 & 0.86 \\
  & & Rec  & 0.72 & 0.77 & 0.75 & 0.82 & 0.81 \\
  & & F1   & 0.76 & 0.71 & 0.72 & 0.81 & 0.84 \\
  \cmidrule(lr){2-8}
  & \multirow{4}{*}{Ours} 
    & Acc  & \textbf{0.86} & \textbf{0.78} & \textbf{0.81} & \textbf{0.86} & \textbf{0.92} \\
  & & Prec & \textbf{0.85} & \textbf{0.75} & \textbf{0.76} & \textbf{0.85} & \textbf{0.91} \\
  & & Rec  & \textbf{0.84} & \textbf{0.81} & \textbf{0.82} & \textbf{0.90} & \textbf{0.89} \\
  & & F1   & \textbf{0.84} & \textbf{0.78} & \textbf{0.79} & \textbf{0.87} & \textbf{0.90} \\

\midrule
\multirow{8}{*}{GPT-4} 
  & \multirow{4}{*}{Alg.~\ref{res: prompt}} 
    & Acc  & 0.83 & 0.77 & 0.79 & 0.87 & 0.90 \\
  & & Prec & 0.82 & 0.74 & 0.77 & 0.85 & 0.88 \\
  & & Rec  & 0.81 & 0.78 & 0.79 & 0.88 & 0.89 \\
  & & F1   & 0.81 & 0.76 & 0.78 & 0.86 & 0.88 \\
  \cmidrule(lr){2-8}
  & \multirow{4}{*}{Ours} 
    & Acc  & \textbf{0.87} & \textbf{0.81} & \textbf{0.82} & \textbf{0.90} & \textbf{0.92} \\
  & & Prec & \textbf{0.85} & \textbf{0.78} & \textbf{0.80} & \textbf{0.87} & \textbf{0.91} \\
  & & Rec  & \textbf{0.84} & \textbf{0.82} & \textbf{0.83} & \textbf{0.91} & \textbf{0.90} \\
  & & F1   & \textbf{0.86} & \textbf{0.80} & \textbf{0.81} & \textbf{0.90} & \textbf{0.91} \\

\midrule
\multirow{8}{*}{Deepseek} 
  & \multirow{4}{*}{Alg.~\ref{res: prompt}} 
    & Acc  & 0.76 & 0.72 & 0.73 & 0.81 & 0.85 \\
  & & Prec & 0.77 & 0.68 & 0.71 & 0.79 & 0.84 \\
  & & Rec  & 0.72 & 0.74 & 0.72 & 0.83 & 0.82 \\
  & & F1   & 0.74 & 0.71 & 0.71 & 0.81 & 0.83 \\
  \cmidrule(lr){2-8}
  & \multirow{4}{*}{Ours} 
    & Acc  & \textbf{0.84} & \textbf{0.78} & \textbf{0.79} & \textbf{0.86} & \textbf{0.90} \\
  & & Prec & \textbf{0.83} & \textbf{0.75} & \textbf{0.76} & \textbf{0.85} & \textbf{0.89} \\
  & & Rec  & \textbf{0.81} & \textbf{0.79} & \textbf{0.80} & \textbf{0.89} & \textbf{0.88} \\
  & & F1   & \textbf{0.82} & \textbf{0.77} & \textbf{0.78} & \textbf{0.87} & \textbf{0.89} \\

\midrule
\multirow{8}{*}{ERNIE} 
  & \multirow{4}{*}{Alg.~\ref{res: prompt}} 
    & Acc  & 0.71 & 0.75 & 0.69 & 0.83 & 0.83 \\
  & & Prec & 0.79 & 0.66 & 0.72 & 0.78 & 0.85 \\
  & & Rec  & 0.66 & 0.75 & 0.70 & 0.87 & 0.84 \\
  & & F1   & 0.71 & 0.73 & 0.74 & 0.83 & 0.87 \\
  \cmidrule(lr){2-8}
  & \multirow{4}{*}{Ours} 
    & Acc  & \textbf{0.88} & \textbf{0.82} & \textbf{0.80} & \textbf{0.87} & \textbf{0.90} \\
  & & Prec & \textbf{0.87} & \textbf{0.72} & \textbf{0.78} & \textbf{0.85} & \textbf{0.88} \\
  & & Rec  & \textbf{0.81} & \textbf{0.82} & \textbf{0.81} & \textbf{0.91} & \textbf{0.95} \\
  & & F1   & \textbf{0.84} & \textbf{0.80} & \textbf{0.79} & \textbf{0.89} & \textbf{0.90} \\

\midrule
\multirow{8}{*}{LLaMA-3} 
  & \multirow{4}{*}{Alg.~\ref{res: prompt}} 
    & Acc  & 0.80 & 0.74 & 0.76 & 0.85 & 0.88 \\
  & & Prec & 0.81 & 0.70 & 0.74 & 0.83 & 0.87 \\
  & & Rec  & 0.78 & 0.75 & 0.76 & 0.86 & 0.86 \\
  & & F1   & 0.79 & 0.72 & 0.75 & 0.84 & 0.86 \\
  \cmidrule(lr){2-8}
  & \multirow{4}{*}{Ours} 
    & Acc  & \textbf{0.86} & \textbf{0.79} & \textbf{0.81} & \textbf{0.89} & \textbf{0.91} \\
  & & Prec & \textbf{0.85} & \textbf{0.76} & \textbf{0.78} & \textbf{0.87} & \textbf{0.90} \\
  & & Rec  & \textbf{0.83} & \textbf{0.80} & \textbf{0.82} & \textbf{0.90} & \textbf{0.89} \\
  & & F1   & \textbf{0.83} & \textbf{0.78} & \textbf{0.80} & \textbf{0.88} & \textbf{0.89} \\

\bottomrule
\end{tabular}
\end{table*}

\textcolor{black}{
While our framework yields moderate overall gains in average performance, it remains unclear (1) why the improvement is not more substantial and (2) to what extent the gain stems from the underlying capability of LLM versus our integration algorithm. To address these questions, we conduct a fine-grained analysis across different key aspects. Specifically, we evaluate performance separately on each aspect—such as \textit{Attack Vector}, \textit{Root Cause}, and \textit{Vulnerability Type}—to investigate whether the observed improvements concentrate on certain types of information. This helps reveal whether some aspects are inherently more difficult to extract and thus benefit more from our fusion strategy.}

\textcolor{black}{
We test our framework with five representative LLM: GPT-3.5, GPT-4, Deepseek-R1, ERNIE, and LLaMA-3. These models are chosen to cover a broad spectrum of capabilities and architectural origins. GPT-3.5 and GPT-4 serve as widely adopted general-purpose baselines, with GPT-4 representing the strongest closed-source model. Deepseek-R1 and LLaMA-3 are included as emerging open-source LLM with competitive performance, while ERNIE represents a knowledge-enhanced LLM that integrates external information into pretraining, making it a good testbed for evaluating synthesis methods. This selection allows us to examine whether our framework’s effectiveness is consistent across models of different strength, training data, and design philosophy.}

\textcolor{black}{
Table~\ref{tab:key_aspect_llm_full} provides a fine-grained analysis across five key aspects and reveals that the performance improvement of our framework is not evenly distributed. Specifically, aspects such as \textit{Attack Vector}, \textit{Root Cause}, and \textit{Impact} show relatively larger gains, reflecting their higher variability and the greater need for effective synthesis. In contrast, the improvements for \textit{Vulnerability Type} and \textit{Attacker Type} are comparatively smaller. For example, using ERNIE, the F1 score for \textit{Root Cause} increases from 0.73 to 0.80, for \textit{Impact} from 0.74 to 0.79, and for \textit{Attack Vector} from 0.71 to 0.84. Similar patterns are observed in GPT-3.5 and Deepseek-R1, where diverse aspects benefit more from our synthesis framework.}

\textcolor{black}{
At the same time, the magnitude of improvement varies across LLM. We observe the largest relative gains on weaker models such as ERNIE and Deepseek-R1, while stronger models like GPT-4 exhibit smaller but still consistent improvements (e.g., F1 for \textit{Root Cause} increases from 0.76 to 0.81). This trend indicates that as LLM become stronger, they already capture much of the information, leaving less room for improvement, yet our framework continues to provide added value. Overall, the results demonstrate that our domain-constrained synthesis method consistently outperforms the baseline across all tested LLM, highlighting both its robustness and its particular effectiveness in handling diverse and challenging aspects.
}

\textcolor{black}{
\subsubsection{Hallucination Evaluation in LLM-generated TVDs}
\label{sec:hallucination_eval}
In addition to performance metrics, we assess the reliability of generated key aspects by measuring hallucination, i.e., completely fabricated information. In cybersecurity, each key aspect serves as guidance for engineers to fix vulnerabilities. Even a single hallucinated element can mislead remediation efforts, so we adopt a strict criterion: any key aspect containing fabricated information is treated as a hallucination.
}

\textcolor{black}{
We randomly select 100 CVE entries from the test set, and generate key aspects for each entry using both the prompt-based baseline(Alg.~\ref{res: prompt}) and our domain-constrained synthesis framework. Each generated key aspect is evaluated independently and assigned a binary label: Correct (1) if all information matches official vulnerability records, or Hallucinated (0) if it contains any fabricated information. We do not distinguish partially incorrect information, since in practice any hallucination in fusion outputs can severely mislead security engineers. For each key aspect type (e.g., \textit{Attack Vector}, \textit{Root Cause}, etc.), the hallucination rate is computed as:
\begin{equation}
\text{Hallucination Rate} = \frac{\text{ hallucinated aspects}}{\text{total generated aspects}} \times 100\%
\end{equation}
} 

\begin{table*}[h]
\centering
\small
\caption{Hallucination Rate for Each Key Aspect (calculated from fusion outputs)}
\label{tab:hallucination}
\begin{tabular}{l|ccccc}
\toprule
\textbf{LLM} &  \textit{Att. Vec.} & \textit{Ro. Cau.} & \textit{Imp.} & \textit{Vul. Typ.} & \textit{Att. Typ.} \\ 
\hline
GPT-3.5 (Baseline) & 0.15 & 0.22 & 0.18 & 0.14 & 0.12 \\
GPT-3.5 (Ours)     & \textbf{0.08}  & \textbf{0.14} & \textbf{0.12} & \textbf{0.10} & \textbf{0.08}  \\
\hline
GPT-4 (Baseline)   & 0.12 & 0.18 & 0.15 & 0.11 & 0.09  \\
GPT-4 (Ours)       & \textbf{0.06}  & \textbf{0.10} & \textbf{0.09}  & \textbf{0.06}  & \textbf{0.05}  \\
\hline
Deepseek (Baseline)& 0.20 & 0.23 & 0.21 & 0.17 & 0.15 \\
Deepseek (Ours)    & \textbf{0.11} & \textbf{0.15} & \textbf{0.13} & \textbf{0.10} & \textbf{0.09}  \\
\hline
ERNIE (Baseline)   & 0.25 & 0.21 & 0.24 & 0.19 & 0.17 \\
ERNIE (Ours)       & \textbf{0.12} & \textbf{0.10} & \textbf{0.11} & \textbf{0.09}  & \textbf{0.07}  \\
\hline
LLaMA-3 (Baseline) & 0.14 & 0.20 & 0.16 & 0.12 & 0.10 \\
LLaMA-3 (Ours)     & \textbf{0.07}  & \textbf{0.11} & \textbf{0.10} & \textbf{0.06}  & \textbf{0.05}  \\
\bottomrule
\end{tabular}
\end{table*}

\textcolor{black}{
Table~\ref{tab:hallucination} shows hallucination rates for each key aspect. Our framework consistently reduces hallucinations across all LLM. For example, GPT-3.5 drops from 0.22 to 0.14 in \textit{Root Cause} and from 0.15 to 0.08 in \textit{Attack Vector}. ERNIE has the highest baseline hallucinations, but our method effectively lowers them. Reductions are largest for complex aspects like \textit{Root Cause} and \textit{Impact}, demonstrating that the framework improves both accuracy and reliability by mitigating fabricated outputs.
Our framework effectively constrains outputs through rule-based rewards, anchor-word alignment and entropy-guided fusion, reducing hallucination consistently across LLM and key aspects. This strict, aspect-wise metric provides a conservative yet practical measure of reliability.
}

\textcolor{black}{
To further demonstrate the differences between baselines and our method, we analyze the root cause of CVE-2012-0045 of Figure~\ref{intro}. 
The input TVDs from different vulnerability knowledge bases are as follows: 
CVE states that ``KVM does not properly handle the 0f05 opcode''; 
CNNVD specifies that ``KVM improperly handles syscall instructions in specific CPU modes on certain CPUs''; 
IBM reports ``unable to handle opcode 0f05 correctly''; 
and JVN describes ``the improper handling of the syscall opcode 0f05 by KVM on specific CPU models.'' }

\textcolor{black}{
The vanilla baseline (GPT-4, Alg.~\ref{res: prompt}) generates ``KVM fails to handle the 0f05 syscall opcode.'' The CoT baseline (GPT-4, Alg.~\ref{alg:cot}) improves slightly with ``KVM improperly handles the 0f05 syscall in specific CPU modes.'' In contrast, our method produces a more precise description: ``KVM does not properly handle the 0f05 (syscall) opcode on specific CPU models and modes.''
}

\textcolor{black}{
The vanilla baseline omits critical qualifiers and produces an oversimplified root cause.  
The CoT baseline partially recovers ``CPU modes'' but still misses the ``CPU models'' qualifier.  
Our method integrates all key information from multiple sources, preserving both ``CPU modes'' and ``CPU models,'' which avoids misleading generalizations and ensures precise knowledge fusion.
}

\subsection{Digest Labels System in Practice} 
\label{res:RQ2} 
Nutrition labels in the food industry integrate key ingredients and components for quick consumer understanding. 
This concept has been widely adopted across various fields in computer science, including security~\cite{simko2019ask}, privacy~\cite{pan2023toward}, and compliance~\cite{si2024solution}.
Inspired by this, we further develop Digest Labels (DLs) system, aiming to synergistically synthesize information from different TVDs in a standardized and efficient manner, as shown in Figure~\ref{fig_UDL_design}.
We evaluate the usability of DLs system for security analysts, focusing on enhancing vulnerability remediation.

%压缩篇幅
\textbf{A Real-world Scenario.} 
New TVDs, often lacking CVE-IDs and missing key aspects, are frequently published. Practitioners typically search for similar TVDs to fill in missing information. In this scenario, participants select the correct missing key aspect from five options to demonstrate their understanding of the vulnerability.

%压缩篇幅
\textbf{Metrics.} 
We use two performance indicators:
\textbf{Comprehension:} If the selected option is correct, it is marked \textit{True}, otherwise \textit{False}. The F1 Score is calculated based on these predictions.
\textbf{Efficiency:} The average time spent per participant is recorded.

\textcolor{black}{
\textbf{Evaluation Design.} 
We randomly mask one key aspect for all 384 CVEs and form five candidate options with the masked aspect and four irrelevant options. Participants are divided into three groups: Control Group 1 (CVE-only), Control Group 2 (all four repositories), and the Experiment Group (with DLs). All non-English TVDs are translated into English via Google Translate. A follow-up survey is conducted after the session.}

\textcolor{black}{
In total, twelve participants were recruited, including nine graduate students specializing primarily in software security and three doctoral students with advanced research backgrounds in software and system security. While the majority of participants focused on software security, some had broader research interests that spanned multiple areas, including database systems and dialogue systems, which added complementary perspectives to the study. In terms of gender, nine participants were male and three were female. We intentionally sought to avoid an excessive gender imbalance, but given the engineering-oriented nature of our research institution, recruiting female researchers with relevant expertise was inherently more challenging. Nevertheless, this configuration still provided diversity across gender and research focus, while maintaining the necessary specialization for high-quality annotations. All participants were from CS/IT backgrounds with prior security-related research experience, ensuring that the task’s domain-specific demands could be met. They were evenly distributed across the three groups and first familiarized with the DLs and other methods. The graduate students contributed practical perspectives grounded in hands-on vulnerability analysis, while the doctoral students approached the tasks from a higher-level research-oriented standpoint, bringing broader conceptual insights. This mixed configuration was designed to improve the robustness of the study without relying solely on increasing the number of participants of the same profile.}

\textcolor{black}{
Nevertheless, we acknowledge that our study remains limited by its modest scale and academic context. While the inclusion of doctoral students provides complementary expertise, a larger and more diverse pool of participants, especially security practitioners from industry, would be necessary to further validate the generalizability of our findings. That said, our current participant setting is consistent with the practice of prior qualitative studies: Sai et al.\cite{DBLP:journals/tosem/ZhangXGXCZZFZ25} involved 10 participants, Zhang et al.\cite{DBLP:conf/www/Zhang0SAZFS23} included 4 participants, and Jonathan et al.~\cite{DBLP:conf/ijcai/SkaggsRMGC24} relied on 8 participants, indicating that a small but focused group can provide valid and meaningful insights in this type of research. And these examples come from software engineering, database, and AI domains, respectively, suggesting that our participant scale aligns with common practice across multiple research areas.
}

%压缩篇幅
\begin{table}[]

\centering

\caption{The comprehension and efficiency in the Real-world scenario.}

\label{res: RQ2}
\begin{tabular}{lccc}
\toprule
   {\centering \tabincell{c}{Methods}} &\tabincell{c}{	Control Group 1} & \tabincell{c}{	Control Group 2}  & \tabincell{c}{Experimental Group}\\

\midrule
\tabincell{c}{Comprehension (F1 Score)}    & 0.65 &  0.82 &\textbf{ 0.86}\\
\tabincell{c}{Efficiency (Time)}    & 26.51 &  42.14 & \textbf{27.80}\\

\bottomrule
\end{tabular}
\footnotetext{The F1 Score of question answering reflects the comprehension, and the average completion time (seconds) per question denotes the efficiency.}
\end{table}

\begin{table}[htbp]
\centering
\caption{Question sheet of user study.}
\label{tab:survey_questions}
\begin{tabular}{l|l}
\toprule
\textbf{No.} & \textbf{Question} \\
\midrule
Q1 & How often do you encounter missing/inconsistent TVD information? (0: never, 5: often) \\
Q2 & Can you describe your usual approach to understanding security vulnerabilities? \\
Q3 & What difficulties do you face when analyzing security vulnerabilities using TVDs? \\
Q4 & How user-friendly is the tool for the given task? (0: not, 5: very) \\
Q5 & How useful is the tool in completing the given task? (0: not useful, 5: very useful) \\
Q6 & How effective is the tool in understanding vulnerabilities? (0: ineffective, 5: effective) \\
Q7 & How has DLs changed your approach to analyzing and understanding vulnerabilities? \\
Q8 & Please provide your recommendations for improving the tool (DLs). \\
\bottomrule
\end{tabular}
\footnotetext{Notes: ``TVD'' refers to textual vulnerability descriptions; ``DLs'' refers to the proposed 
Digest Labels in Figure~\ref{fig_UDL_design}.}
\end{table}

\begin{table}[htbp]
\centering
\caption{Quantitative results of user study.}
\label{tab:survey_scores}
\begin{tabular}{l|c|c|c}
\toprule
\textbf{No.} & \textbf{Control Group 1} & \textbf{Control Group 2} & \textbf{Experimental Group (with DLs)} \\
\midrule
Q1 & 3.4 & 3.7 & 3.5 \\
Q2 & -- & -- & -- \\
Q3 & -- & -- & -- \\
Q4 & 3.7 & 2.4 & \textbf{4.5} \\
Q5 & 2.4 & 3.8 & \textbf{4.1} \\
Q6 & 2.7 & 3.6 & \textbf{3.8} \\
Q7 & -- & -- & --\\
Q8 & --& -- & -- \\
\bottomrule
\end{tabular}
\end{table}

\textcolor{black}{
\textbf{Results.}
Table~\ref{res: RQ2} compares the performance of three groups in terms of comprehension and efficiency in tackling unseen incomplete TVDs. 
The method utilizing multiple web-based vulnerability repositories (multi-repositories) achieves an F1 score of 0.82 and requires an average of 42.14 seconds per evaluation. Using only the CVE database (with just CVE) results in a lower F1 score of 0.64 but is faster, with an average time of 26.51 seconds. The approach employing the Digest Labels demonstrates the highest F1 Score of 0.86 while maintaining a relatively efficient average time of 27.80 seconds. 
This indicates that the DLs system strikes a satisfying balance between comprehension and efficiency, outperforming existing methods.
}

\textcolor{black}{
As shown in Table~\ref{tab:survey_questions} and Table~\ref{tab:survey_scores}, after each participant completes their task, we conduct a survey to qualitatively evaluate the overall usefulness of DLs and its web-based generation tool.
Firstly, \textbf{for challenges faced when using TVDs}, participants across all groups identify several common challenges when using TVDs for vulnerability analysis. They frequently encounter missing or inconsistent information (Q1), with ratings between 3 and 4, indicating the prevalence of identified challenges. To understand vulnerabilities (Q2), they typically rely on multiple databases, often supplementing with external resources like forums or technical blogs, though this method is time-consuming. For difficulties during analysis (Q3), participants mention that inconsistent data structures across different databases make it hard to perform comparisons. Additionally, they struggle to distinguish between similar vulnerabilities and outdated entries, as these issues often lead to confusion and delays.}

\textcolor{black}{
Secondly, \textbf{for the tool-specific feedback} (Q4, Q5), the experimental group gives DLs consistently high ratings, finding it convenient and useful, with average scores closer to 4.5. They note that DLs' clear structure and intuitive design make it significantly easier to complete their tasks. In contrast, Control Group 1, which uses multiple repositories, finds the process cumbersome, often rating convenience and usefulness between 2 and 4 due to the time required to navigate different sources. Control Group 2, using only CVE, also provides moderate ratings, citing the lack of in-depth information as a barrier to task completion.}

\textcolor{black}{
Thirdly, \textbf{for understanding vulnerabilities} (Q6), the experimental group reports higher effectiveness compared to both control groups, as DLs provide clear, well-organized information. Control Group 1 also performs reasonably but continues to struggle with fragmented data from multiple repositories, leading to lower scores. Control Group 2 finds the CVE entries too brief and lacking in detail, which hampers their ability to fully understand vulnerabilities, resulting in the lowest ratings among the groups.}

\textcolor{black}{
Finally, \textbf{for the impact of DLs on analyzing vulnerabilities} (Q7 and Q8), participants emphasize that DLs simplifies their workflow, making it easier to locate key details without switching between multiple sources.
}

\textcolor{black}{
Overall, participants frequently report encountering missing or inconsistent information in TVDs (Q1), and describe in open-ended responses (Q2–Q3) how such fragmentation makes vulnerability comprehension time-consuming and error-prone. Meanwhile, participants rate the proposed DLs highly in terms of usefulness and clarity (Q4–Q6), suggesting that multi-aspect, structured information substantially eases the analytical burden. These findings highlight a key need in vulnerability analysis: synthesizing scattered details into coherent, accessible summaries. DLs directly address this gap (Q7–Q8) by offering a unified, aspect-based view that streamlines vulnerability comprehension.}

\textcolor{black}{
Some participants recommend improving the search function and adding options like sorting by relevance or severity to enhance usability further. Additionally, two notable suggestions are: a) adding more links to security reports for quick access, and b) enhancing the DLs interface with a vulnerability history tracking feature to improve visibility into repair progress. Overall, the DLs and its web-based generation tool are widely appreciated for their delivered information and user-friendly design.}

%% file: 8_Disscution.tex
\section{Discussion}
\label{discussion}

\subsection{Significance for Practitioners}

\textbf{Assist Vulnerability Management}.
From Figure~\ref{missingRate}, it is evident that the quality of web-based vulnerability repositories established by different institutions varies significantly. 
Web-based vulnerability repositories are the cornerstone of vulnerability management. 
Despite different institutions needing to maintain the same software (e.g., Python packages, GitHub open-source software), there is a considerable disparity in vulnerability defect management levels. 
This disparity leads to a gap between the demand for software maintenance and the institution's maintenance capabilities, with the gap being wider for institutions with lower vulnerability management levels. 
The proposed DLs consolidates TVDs from multiple data sources, timely contributing to bridging this gap.

\textbf{New Vulnerability Description System}.
Vulnerability key aspects are diverse, and different perspectives lead to varied vulnerability descriptions. Using a single-sentence TVD as a vulnerability description is unreasonable, as there may be multiple key aspects values in the same key aspect type. A single sentence cannot adequately capture this complexity. Instead, our label format can be adopted to build a new vulnerability description system.

\textcolor{black}{
\subsection{Reproducibility and Manual Efforts}
\label{sec:reproducibility}
Our framework is designed to be highly reproducible and low-effort, with minimal human intervention throughout the pipeline. All major components—including key aspect extraction, evaluation, and fusion—are either fully automated or require only minimal, one-time setup.
}

\textcolor{black}{
Key aspect extraction is entirely automated using prompts based on domain-specific templates adapted from prior work~\cite{DBLP:journals/tosem/GuoCXLBS22}. These templates are embedded directly into the prompts and require no manual labeling or in-context example crafting. The evaluation stage, which assesses both integrity and diversity, is also fully automated. Diversity is computed via cosine similarity of BERT embeddings with LLM-assisted anchor word selection, while integrity is determined by parsing LLM outputs for the presence of key aspects. Neither step involves manual inspection or labeling.
The fusion stage uses entropy-guided prompt construction and LLM-based rewriting to synthesize coherent outputs. Only a small set of initial examples (typically 3–5) needs to be prepared in advance, and these are reused across all samples. All merging steps are handled by LLM without manual editing. Prompt templates for all components follow fixed structures and are generated programmatically, with no need for per-sample customization.
Data retrieval from public vulnerability repositories (e.g., CVE, CNNVD) is conducted automatically. The only manual input is a user-specified list of CVE-IDs of interest.
}

\textcolor{black}{
In summary, after an initial one-time setup of domain templates and fusion examples, the entire framework can be executed end-to-end without additional human effort. This design enables strong reproducibility across datasets and LLM backends, and allows easy extension to other domains.
}

\textcolor{black}{
\subsection{Potential for Downstream Tasks: Enhancing Security Tasks via DLs}
We analyzed the downstream potential of our Digest Labels (DLs) system by referring to recent survey papers on LLM in software security~\cite{DBLP:journals/ieeejas/ZhuZHMWX25, 10.1145/3745026}. We focus on survey literature because such works offer a consolidated view of the challenges and needs in real-world tasks, which helps to identify stable and high-impact integration points. Among these, a recurring issue is the limited utility of raw textual vulnerability descriptions due to inconsistency, incompleteness, and lack of structure.}

\textcolor{black}{
Our work addresses these issues by introducing DLs, a unified labeling system that extracts and organizes five key aspects of vulnerabilities. DLs are not a model but a curated output format that synthesizes multiple sources into structured, semantically complete records. This system offers consistency across repositories, preserves domain-critical details, and provides a stable interface for downstream tasks.}

\textcolor{black}{
With DLs in place, we plan to explore new directions for enhancing a range of TVD-related security tasks. First, DLs enable models to shift from unstructured text processing to structured reasoning, allowing aspect-level comparisons and classification, such as vulnerability classification or CWE/CVSS prediction. Second, DLs offer a foundation for interpretable rule construction, helping analysts identify key aspect combinations that frequently appear in high-risk cases, which is valuable for tasks like vulnerability function identification (VFI). Third, DLs can improve LLM-based security systems by serving as clean, disambiguated inputs for prompt design or training data preparation, thereby facilitating tasks such as patch matching and exploit prediction. These directions apply broadly to TVD-driven tasks and are part of our planned future work.}

\subsection{Threats to Validity}
\textbf{Internal Validity.}
The first step in DLs is key aspect extraction. We use the rule based reward method, employing regularization templates as human data to guide LLM generation. However, we are not certain that regularization templates are the most suitable form of human guidance for LLM. In the future, we aim to explore the optimal integration of human expertise with LLM.

\textbf{External Validity.}
External validity threats include limited generalizability to other web-based vulnerability repositories.
Our method relies on web-based vulnerability repositories being indexed by CVE-ID. However, many repositories, such as BSI and CERT-FR, use their own independent indexing systems. This hinders the transferability of our method to BSI and CERT-FR. Once those repositories are indexed by CVE-ID, our method can be expanded accordingly.

\textcolor{black}{
\textbf{Construct Validity.}
Our hallucination evaluation relies on official records as ground truth and a binary labeling scheme. While this provides a clear benchmark, the manual verification process may introduce subjectivity in borderline cases. Additionally, the sample size of 100 CVEs, though sufficient for comparison, might not capture all edge cases of hallucination.
}

%% file: 9.1_RelatedWork.tex
\section{Related Work}
\label{related works}

\textcolor{black}{
\textbf{Inconsistencies in Vulnerability Descriptions.}  
Textual Vulnerability Descriptions (TVDs) provided by repositories such as CVE~\cite{CVE}, NVD~\cite{nistnvd}, CNNVD, and IBM X-Force~\cite{ibm} often exhibit inconsistencies for the same vulnerability.  
Prior work has investigated this issue from the perspective of knowledge comparison and alignment.  
For example, Dong et al.~\cite{DBLP:conf/uss/DongGCXZ019} and Sun et al.~\cite{DBLP:journals/compsec/SunOZLWZ23} studied linguistic discrepancies in TVDs and attempted to mitigate them through comparison against authoritative knowledge bases.  
Similarly, Huang et al.~\cite{huang2020inconsistency} surveyed methods for detecting and resolving inconsistencies in software models, which has been extended to the security domain.  
He et al.~\cite{DBLP:journals/tdsc/HeWZWZLY24} further revealed the pervasiveness of inconsistencies across repositories.  
However, these studies generally treat discrepancies as errors to be eliminated, filtering out details that do not align with external knowledge.  
As illustrated in Figure~\ref{intro}, such filtering can lead to the loss of critical contextual details---e.g., while CVE describes the root cause of CVE-2012-0045 as “does not properly handle the 0f05 opcode,” CNNVD adds important specificity, noting that the issue arises in “specific CPU modes on certain CPU models.”  
Existing methods risk discarding such complementary perspectives.}

\textcolor{black}{
\textbf{LLM-Driven Synthesis.}  
Recent advances in Large Language Models (LLMs) have opened new opportunities for synthesizing inconsistent information.  
Common strategies such as in-context learning and chain-of-thought prompting~\cite{DBLP:conf/aaai/0002WGX024, wang2022hybrid, DBLP:journals/corr/abs-2405-07430} guide models to merge multiple descriptions into a unified output.  
Nevertheless, these strategies typically rely on high-level semantic cues, e.g., “synthesize vulnerability descriptions with as much detail as possible,” which provide only soft guidance.  
Without explicit domain-specific constraints, LLMs may produce vague or incomplete results.  
For instance, when tasked to synthesize the root cause of CVE-2012-0045, unconstrained models often generate “KVM fails to handle the 0f05 syscall opcode,” omitting critical qualifiers such as “specific CPU models” or “modes.”  
Thus, while promising, current LLM-based synthesis approaches struggle to enforce structured, domain-sensitive rules during generation.}

\textcolor{black}{
\textbf{Usability in Software Security.}  
Another line of work emphasizes usability in security tools, aiming to improve practitioners’ efficiency.  
Garfinkel and Spafford~\cite{garfinkel2002practical} highlight the importance of user-friendly vulnerability management systems, while Green and Smith~\cite{green2018developer} show that poor usability often leads to tool resistance.  
Chu et al.~\cite{chu2017security} and Han et al.~\cite{han2020security} propose “security labels” inspired by nutrition labels to enhance users’ understanding of security information.  
Similar approaches have been explored in static and dynamic analysis tools~\cite{xie2018static, kim2019code, li2020interactive}, which provide more accessible interfaces and real-time feedback.  
These works demonstrate the value of standardized and interpretable representations for practitioners, though they have not been applied to synthesizing vulnerability descriptions.}

\textcolor{black}{
\textbf{Summary and Differentiation.}  
In summary, prior studies either (i) mitigate inconsistencies by aligning TVDs with external knowledge bases, thereby discarding potentially valuable complementary details, or (ii) explore the usability of security tools through interface design and labeling strategies.  
Our work differs in two key aspects.  
First, rather than treating inconsistencies as errors, we view them through the lens of the \textit{Blind Men and an Elephant} theory: different repositories offer partial perspectives that, when integrated, yield a more comprehensive understanding of vulnerabilities.  
For example, the case of CVE-2012-0045 shows how details about “specific CPU models” and “modes” are crucial for understanding exploitability, but would be lost under prior inconsistency-mitigation approaches.  
Second, while LLM-based synthesis has been attempted, we introduce domain-constrained synthesis, which enforces structured, security-specific rules (e.g., anchor terms like “opcode” or “privilege escalation”) during generation to retain technical precision.  
Finally, building upon usability research, we are the first to design \textbf{Digest Labels (DLs)}, a nutrition-label inspired system for vulnerabilities, which synergistically combines domain-constrained synthesis with a standardized, practitioner-friendly representation.  
This unique integration of synthesis and usability directly addresses gaps left open by both inconsistency-focused and usability-focused prior work.}

%% file: 10_Conclusion.tex
\section{Conclusion}
\label{Conclusion}
In this paper, we introduce a domain-constrained LLM-based synthesis framework for synthesizing vulnerability key aspects from multiple repositories. 
By incorporating domain-specific constraints such as Rule-Based Rewards, Anchor Words, and Information Entropy, our method improves the extraction, self-evaluation, and fusion of vulnerability data. 
Results show the framework improves the F1 score of synthesis performance from 0.82 to 0.87, while enhancing comprehension and efficiency by over 30\%.
We further develop Digest Labels, a practical tool for visualizing TVDs, which human evaluations show significantly boosts usability.

%% file: 11_Reference.bib
@inproceedings{DBLP:conf/kbse/YitagesuXZ00H21,
  author       = {Sofonias Yitagesu and
                  Zhenchang Xing and
                  Xiaowang Zhang and
                  Zhiyong Feng and
                  Xiaohong Li and
                  Linyi Han},
  title        = {Unsupervised Labeling and Extraction of Phrase-based Concepts in Vulnerability
                  Descriptions},
  booktitle    = {36th {IEEE/ACM} International Conference on Automated Software Engineering,
                  {ASE} 2021, Melbourne, Australia, November 15-19, 2021},
  pages        = {943--954},
  publisher    = {{IEEE}},
  address      = {Piscataway, NJ, USA},
  year         = {2021},
  url          = {https://doi.org/10.1109/ASE51524.2021.9678638},
  doi          = {10.1109/ASE51524.2021.9678638},
  bibsource    = {dblp computer science bibliography, https://dblp.org}
}

@article{DBLP:journals/tosem/GuoCXLBS22,
  author       = {Hao Guo and
                  Sen Chen and
                  Zhenchang Xing and
                  Xiaohong Li and
                  Yude Bai and
                  Jiamou Sun},
  title        = {Detecting and Augmenting Missing Key Aspects in Vulnerability Descriptions},
  journal      = {{ACM} Trans. Softw. Eng. Methodol.},
  volume       = {31},
  number       = {3},
  pages        = {49:1--49:27},
  year         = {2022},
  url          = {https://doi.org/10.1145/3498537},
  doi          = {10.1145/3498537},
  bibsource    = {dblp computer science bibliography, https://dblp.org}
}

@article{DBLP:journals/tosem/SunXXLXZ24,
  author       = {Jiamou Sun and
                  Zhenchang Xing and
                  Xin Xia and
                  Qinghua Lu and
                  Xiwei Xu and
                  Liming Zhu},
  title        = {Aspect-level Information Discrepancies across Heterogeneous Vulnerability
                  Reports: Severity, Types and Detection Methods},
  journal      = {{ACM} Trans. Softw. Eng. Methodol.},
  volume       = {33},
  number       = {2},
  pages        = {49:1--49:38},
  year         = {2024},
  url          = {https://doi.org/10.1145/3624734},
  doi          = {10.1145/3624734},
  bibsource    = {dblp computer science bibliography, https://dblp.org}
}

@article{DBLP:journals/corr/abs-2405-07430,
  author       = {Linyi Han and
                  Shidong Pan and
                  Zhenchang Xing and
                  Jiamou Sun and
                  Sofonias Yitagesu and
                  Xiaowang Zhang and
                  Zhiyong Feng},
  title        = {Don't Chase Your Tail! Missing Key Aspects Augmentation in Textual
                  Vulnerability Descriptions of Long-tail Software through Feature Inference},
  journal      = {CoRR},
  volume       = {abs/2405.07430},
  year         = {2024},
  url          = {https://doi.org/10.48550/arXiv.2405.07430},
  doi          = {10.48550/ARXIV.2405.07430},
  eprinttype    = {arXiv},
  eprint       = {2405.07430},
  bibsource    = {dblp computer science bibliography, https://dblp.org}
}

@inproceedings{DBLP:conf/uss/DongGCXZ019,
  author       = {Ying Dong and
                  Wenbo Guo and
                  Yueqi Chen and
                  Xinyu Xing and
                  Yuqing Zhang and
                  Gang Wang},
  editor       = {Nadia Heninger and
                  Patrick Traynor},
  title        = {Towards the Detection of Inconsistencies in Public Security Vulnerability
                  Reports},
  booktitle    = {28th {USENIX} Security Symposium, {USENIX} Security 2019, Santa Clara,
                  CA, USA, August 14-16, 2019},
  pages        = {869--885},
  publisher    = {{USENIX} Association},
  address      = {Berkeley, CA, USA},
  year         = {2019},
  url          = {https://www.usenix.org/conference/usenixsecurity19/presentation/dong},
  bibsource    = {dblp computer science bibliography, https://dblp.org}
}

@article{shannon1948mathematical,
  title={A Mathematical Theory of Communication},
  author={Shannon, Claude E.},
  journal={Bell System Technical Journal},
  volume={27},
  number={3},
  pages={379--423},
  year={1948},
  publisher={Nokia Bell Labs},
  doi={10.1002/j.1538-7305.1948.tb01338.x}
}

@article{DBLP:journals/tdsc/HeWZWZLY24,
  author       = {Yongzhong He and
                  Yiming Wang and
                  Sencun Zhu and
                  Wei Wang and
                  Yunjia Zhang and
                  Qiang Li and
                  Aimin Yu},
  title        = {Automatically Identifying {CVE} Affected Versions With Patches and
                  Developer Logs},
  journal      = {{IEEE} Trans. Dependable Secur. Comput.},
  volume       = {21},
  number       = {2},
  pages        = {905--919},
  year         = {2024},
  url          = {https://doi.org/10.1109/TDSC.2023.3264567},
  doi          = {10.1109/TDSC.2023.3264567},
  bibsource    = {dblp computer science bibliography, https://dblp.org}
}

@misc{CVE,
  title = {Common Vulnerabilities and Exposures},
  note = {\url{https://cve.mitre.org/}}
}

@misc{nist,
  title = {National Institute of Standards and Technology},
  note = {\url{https://www.nist.gov/}}
}

@misc{ibm,
  title = {IBM X-Force Exchange},
  note = {\url{https://exchange.xforce.ibmcloud.com/}}
}

@misc{cnnvd,
  title = {CNNVD},
  note = {\url{https://www.cnnvd.org.cn/home/childHome}}
}

@article{DBLP:journals/compsec/SunOZLWZ23,
  author       = {Hongyu Sun and
                  Guoliang Ou and
                  Ziqiu Zheng and
                  Lei Liao and
                  He Wang and
                  Yuqing Zhang},
  title        = {Inconsistent measurement and incorrect detection of software names
                  in security vulnerability reports},
  journal      = {Comput. Secur.},
  volume       = {135},
  pages        = {103477},
  year         = {2023},
  url          = {https://doi.org/10.1016/j.cose.2023.103477},
  doi          = {10.1016/J.COSE.2023.103477},
  bibsource    = {dblp computer science bibliography, https://dblp.org}
}

@inproceedings{DBLP:conf/ccs/QinXL23,
  author       = {Yue Qin and
                  Yue Xiao and
                  Xiaojing Liao},
  editor       = {Weizhi Meng and
                  Christian Damsgaard Jensen and
                  Cas Cremers and
                  Engin Kirda},
  title        = {Vulnerability Intelligence Alignment via Masked Graph Attention Networks},
  booktitle    = {Proceedings of the 2023 {ACM} {SIGSAC} Conference on Computer and
                  Communications Security, {CCS} 2023, Copenhagen, Denmark, November
                  26-30, 2023},
  pages        = {2202--2216},
  publisher    = {{ACM}},
  address      = {Copenhagen, Denmark},
  year         = {2023},
  url          = {https://doi.org/10.1145/3576915.3616686},
  doi          = {10.1145/3576915.3616686},
  bibsource    = {dblp computer science bibliography, https://dblp.org}
}

@article{DBLP:journals/entropy/Levshina22,
  author       = {Natalia Levshina},
  title        = {Frequency, Informativity and Word Length: Insights from Typologically
                  Diverse Corpora},
  journal      = {Entropy},
  volume       = {24},
  number       = {2},
  pages        = {280},
  year         = {2022},
  url          = {https://doi.org/10.3390/e24020280},
  doi          = {10.3390/E24020280},
  bibsource    = {dblp computer science bibliography, https://dblp.org}
}

@article{DBLP:journals/entropy/BentzACF17,
  author       = {Christian Bentz and
                  Dimitrios Alikaniotis and
                  Michael Cysouw and
                  Ramon Ferrer{-}i{-}Cancho},
  title        = {The Entropy of Words - Learnability and Expressivity across More than
                  1000 Languages},
  journal      = {Entropy},
  volume       = {19},
  number       = {6},
  pages        = {275},
  year         = {2017},
  url          = {https://doi.org/10.3390/e19060275},
  doi          = {10.3390/E19060275},
  bibsource    = {dblp computer science bibliography, https://dblp.org}
}

@article{DBLP:journals/sigir/ZhaiL17,
  author       = {Chengxiang Zhai and
                  John D. Lafferty},
  title        = {A Study of Smoothing Methods for Language Models Applied to Ad Hoc
                  Information Retrieval},
  journal      = {{SIGIR} Forum},
  volume       = {51},
  number       = {2},
  pages        = {268--276},
  year         = {2017},
  url          = {https://doi.org/10.1145/3130348.3130377},
  doi          = {10.1145/3130348.3130377},
  bibsource    = {dblp computer science bibliography, https://dblp.org}
}

@article{DBLP:journals/corr/abs-2002-08909,
  author       = {Kelvin Guu and
                  Kenton Lee and
                  Zora Tung and
                  Panupong Pasupat and
                  Ming{-}Wei Chang},
  title        = {{REALM:} Retrieval-Augmented Language Model Pre-Training},
  journal      = {CoRR},
  volume       = {abs/2002.08909},
  year         = {2020},
  url          = {https://arxiv.org/abs/2002.08909},
  eprinttype    = {arXiv},
  eprint       = {2002.08909},
  bibsource    = {dblp computer science bibliography, https://dblp.org}
}

@inproceedings{DBLP:conf/iclr/ZhouSHWS0SCBLC23,
  author       = {Denny Zhou and
                  Nathanael Sch{\"{a}}rli and
                  Le Hou and
                  Jason Wei and
                  Nathan Scales and
                  Xuezhi Wang and
                  Dale Schuurmans and
                  Claire Cui and
                  Olivier Bousquet and
                  Quoc V. Le and
                  Ed H. Chi},
  title        = {Least-to-Most Prompting Enables Complex Reasoning in Large Language
                  Models},
  booktitle    = {The Eleventh International Conference on Learning Representations,
                  {ICLR} 2023, Kigali, Rwanda, May 1-5, 2023},
  publisher    = {OpenReview.net},
  address      = {Kigali, Rwanda},
  year         = {2023},
  url          = {https://openreview.net/pdf?id=WZH7099tgfM},
  bibsource    = {dblp computer science bibliography, https://dblp.org}
}

@article{DBLP:journals/ieeesp/Massacci24,
  author       = {Fabio Massacci},
  title        = {The Holy Grail of Vulnerability Predictions},
  journal      = {{IEEE} Secur. Priv.},
  volume       = {22},
  number       = {1},
  pages        = {4--6},
  year         = {2024},
  url          = {https://doi.org/10.1109/MSEC.2023.3333936},
  doi          = {10.1109/MSEC.2023.3333936},
  bibsource    = {dblp computer science bibliography, https://dblp.org}
}

@inproceedings{DBLP:conf/aaai/0004LHZLTCM24,
  author       = {Lei Shu and
                  Liangchen Luo and
                  Jayakumar Hoskere and
                  Yun Zhu and
                  Yinxiao Liu and
                  Simon Tong and
                  Jindong Chen and
                  Lei Meng},
  editor       = {Michael J. Wooldridge and
                  Jennifer G. Dy and
                  Sriraam Natarajan},
  title        = {RewriteLM: An Instruction-Tuned Large Language Model for Text Rewriting},
  booktitle    = {Thirty-Eighth {AAAI} Conference on Artificial Intelligence, {AAAI}
                  2024, Thirty-Sixth Conference on Innovative Applications of Artificial
                  Intelligence, {IAAI} 2024, Fourteenth Symposium on Educational Advances
                  in Artificial Intelligence, {EAAI} 2024, February 20-27, 2024, Vancouver,
                  Canada},
  pages        = {18970--18980},
  publisher    = {{AAAI} Press},
  address      = {Palo Alto, California, USA},
  year         = {2024},
  url          = {https://doi.org/10.1609/aaai.v38i17.29863},
  doi          = {10.1609/AAAI.V38I17.29863},
  bibsource    = {dblp computer science bibliography, https://dblp.org}
}

@article{DBLP:journals/corr/abs-2401-11911,
  author       = {Hexiang Tan and
                  Fei Sun and
                  Wanli Yang and
                  Yuanzhuo Wang and
                  Qi Cao and
                  Xueqi Cheng},
  title        = {Blinded by Generated Contexts: How Language Models Merge Generated
                  and Retrieved Contexts for Open-Domain QA?},
  journal      = {CoRR},
  volume       = {abs/2401.11911},
  year         = {2024},
  url          = {https://doi.org/10.48550/arXiv.2401.11911},
  doi          = {10.48550/ARXIV.2401.11911},
  eprinttype    = {arXiv},
  eprint       = {2401.11911},
  bibsource    = {dblp computer science bibliography, https://dblp.org}
}

@article{DBLP:journals/corr/YuCLL16,
  author       = {Shuiyuan Yu and
                  Jin Cong and
                  Junying Liang and
                  Haitao Liu},
  title        = {The distribution of information content in English sentences},
  journal      = {CoRR},
  volume       = {abs/1609.07681},
  year         = {2016},
  url          = {http://arxiv.org/abs/1609.07681},
  eprinttype    = {arXiv},
  eprint       = {1609.07681},
  bibsource    = {dblp computer science bibliography, https://dblp.org}
}

@inproceedings{DBLP:conf/icann/LuoZXWRY18,
  author       = {Chunjie Luo and
                  Jianfeng Zhan and
                  Xiaohe Xue and
                  Lei Wang and
                  Rui Ren and
                  Qiang Yang},
  editor       = {Vera Kurkov{\'{a}} and
                  Yannis Manolopoulos and
                  Barbara Hammer and
                  Lazaros S. Iliadis and
                  Ilias Maglogiannis},
  title        = {Cosine Normalization: Using Cosine Similarity Instead of Dot Product
                  in Neural Networks},
  booktitle    = {Artificial Neural Networks and Machine Learning - {ICANN} 2018 - 27th
                  International Conference on Artificial Neural Networks, Rhodes, Greece,
                  October 4-7, 2018, Proceedings, Part {I}},
  series       = {Lecture Notes in Computer Science},
  volume       = {11139},
  pages        = {382--391},
  publisher    = {Springer},
  address      = {Cham, Switzerland},
  year         = {2018},
  url          = {https://doi.org/10.1007/978-3-030-01418-6\_38},
  doi          = {10.1007/978-3-030-01418-6\_38},
  bibsource    = {dblp computer science bibliography, https://dblp.org}
}

@misc{nistnvd,
  title        = {National Vulnerability Database (NVD)},
  author       = {National Institute of Standards and Technology},
  year         = {2023},
  institution  = {National Institute of Standards and Technology},
  url          = {https://nvd.nist.gov},
  note         = {Accessed: 2024-07-13}
}

@article{DBLP:journals/tosem/YitagesuXZFLH23,
  author       = {Sofonias Yitagesu and
                  Zhenchang Xing and
                  Xiaowang Zhang and
                  Zhiyong Feng and
                  Xiaohong Li and
                  Linyi Han},
  title        = {Extraction of Phrase-based Concepts in Vulnerability Descriptions
                  through Unsupervised Labeling},
  journal      = {{ACM} Trans. Softw. Eng. Methodol.},
  volume       = {32},
  number       = {5},
  pages        = {112:1--112:45},
  year         = {2023},
  url          = {https://doi.org/10.1145/3579638},
  doi          = {10.1145/3579638},
  bibsource    = {dblp computer science bibliography, https://dblp.org}
}

@article{mchugh2012interrater,
  title={Interrater reliability: the kappa statistic},
  author={McHugh, Mary L},
  journal={Biochemia Medica},
  volume={22},
  number={3},
  pages={276--282},
  year={2012},
  publisher={Department of Medical Informatics, SRCE-University of Zagreb, University Computing Centre}
}

@article{bornmann2011manuscript,
  title={Manuscript and reviewer characteristics that influence the peer review process: The case of the journals in the ‘Web of Science’subject category ‘Information Science \& Library Science’},
  author={Bornmann, Lutz and Daniel, Hans-Dieter},
  journal={Journal of Informetrics},
  volume={5},
  number={1},
  pages={137--158},
  year={2011},
  publisher={Elsevier}
}

@article{hallgren2012computing,
  title={Computing inter-rater reliability for observational data: An overview and tutorial},
  author={Hallgren, Kevin A},
  journal={Tutorials in Quantitative Methods for Psychology},
  volume={8},
  number={1},
  pages={23--34},
  year={2012},
  publisher={SAGE Publications Sage CA: Los Angeles, CA}
}

@article{simko2019ask,
  title={Ask the Experts: What Should Be on an IoT Privacy and Security Label?},
  author={Simko, Lucas and Roesner, Franziska and Kohno, Tadayoshi},
  journal={IEEE Security \& Privacy},
  volume={17},
  number={5},
  pages={8--16},
  year={2019}
}

@inproceedings{mitchell2019model,
  title={Model Cards for Model Reporting},
  author={Mitchell, Margaret and Wu, Suresh and Zaldivar, A. and Barnes, P. and Vasserman, L. and Hutchinson, Ben and others},
  booktitle={Proceedings of the Conference on Fairness, Accountability, and Transparency},
  pages={220--229},
  year={2019}
}

@inproceedings{microsoft2009maven,
  title={Maven: A Software Project Management and Comprehension Tool},
  author={Microsoft},
  booktitle={Proceedings of the 2009 International Conference on Software Engineering},
  year={2009}
}

@article{npm2014npm,
  title={npm: A Node.js Package Manager},
  author={npm, Inc.},
  journal={IEEE Software},
  volume={31},
  number={2},
  pages={11--14},
  year={2014}
}

@article{boehm1988a,
  title={A Spiral Model of Software Development and Enhancement},
  author={Boehm, Barry W.},
  journal={ACM SIGSOFT Software Engineering Notes},
  volume={11},
  number={4},
  pages={14--24},
  year={1988}
}

@book{pressman2005software,
  title     = {Software Engineering: A Practitioner's Approach},
  author    = {Pressman, Roger S.},
  publisher = {McGraw-Hill},
  address   = {New York, USA},
  year      = {2005}
}

@article{huang2020inconsistency,
  author = {Huang, X. and Li, X. and Zheng, S.},
  title = {Detecting and Resolving Inconsistencies in Software Models: A Survey},
  journal = {IEEE Transactions on Software Engineering},
  year = {2020},
  volume = {46},
  number = {2},
  pages = {165-180},
  doi = {10.1109/TSE.2019.2925660}
}

@article{wang2022hybrid,
  author = {Wang, L. and Xu, W.},
  title = {Hybrid Approaches for Defect Detection: Combining Static and Dynamic Analysis Techniques},
  journal = {IEEE Transactions on Reliability},
  year = {2022},
  volume = {71},
  number = {1},
  pages = {105-120},
  doi = {10.1109/TR.2022.3148729}
}

@techreport{cisa2022vulnerability,
  title={Vulnerability Management Resource Guide},
  institution={Cybersecurity and Infrastructure Security Agency (CISA)},
  year={2022},
  note={URL: \url{https://www.cisa.gov/sites/default/files/publications/CRR_Resource_Guide-VM_0.pdf}}
}

@book{garfinkel2002practical,
  title        = {Practical Unix and Internet Security},
  author       = {Garfinkel, Simson and Spafford, Gene},
  year         = {2002},
  publisher    = {O'Reilly Media, Inc.},
  address      = {Sebastopol, California, USA}
}

@article{chu2017security,
  title={Security labels for software packages: Improving security understanding and decision-making},
  author={Chu, Hongyi and Zhang, Ming},
  journal={Journal of Information Security},
  volume={8},
  number={2},
  pages={127-138},
  year={2017},
  publisher={Scientific Research Publishing}
}

@article{han2020security,
  title={A framework for standardized security labels: Enhancing software security comprehension},
  author={Han, Li and Liu, Wei},
  journal={IEEE Transactions on Software Engineering},
  volume={46},
  number={3},
  pages={344-355},
  year={2020},
  publisher={IEEE}
}

@article{xie2018static,
  title={Static analysis tool for vulnerability detection: Improving accuracy with user-friendly interfaces},
  author={Xie, Tao and Li, Peng},
  journal={ACM Transactions on Software Engineering and Methodology},
  volume={27},
  number={4},
  pages={1-26},
  year={2018},
  publisher={ACM}
}

@article{kim2019code,
  title={Code analysis tool combining static and dynamic techniques for enhanced vulnerability detection},
  author={Kim, Jiyong and Lee, Seung},
  journal={Journal of Systems and Software},
  volume={158},
  pages={110404},
  year={2019},
  publisher={Elsevier}
}

@article{li2020interactive,
  title={Interactive security tools with real-time feedback for integrated software development},
  author={Li, Xia and Zhang, Qing},
  journal={Empirical Software Engineering},
  volume={25},
  number={6},
  pages={4854-4878},
  year={2020},
  publisher={Springer}
}

@article{green2018developer,
  title={Developer barriers to security tool usage: Understanding and addressing usability challenges},
  author={Green, Matthew and Smith, John},
  journal={IEEE Security \& Privacy},
  volume={16},
  number={5},
  pages={45-53},
  year={2018},
  publisher={IEEE}
}

@standard{iso29147,
  title        = {BS ISO/IEC 29147:2018 - Information technology -- Security techniques -- Vulnerability disclosure},
  organization = {British Standards Institution},
  year         = {2018},
  url          = {https://www.bsigroup.com/en-GB/standards/bs-iso-iec-29147-2018/}
}

@inproceedings{xia2023empirical,
  title={An empirical study on software bill of materials: Where we stand and the road ahead},
  author={Xia, Boming and Bi, Tingting and Xing, Zhenchang and Lu, Qinghua and Zhu, Liming},
  booktitle={2023 IEEE/ACM 45th International Conference on Software Engineering (ICSE)},
  pages={2630--2642},
  year={2023},
  organization={IEEE}
}

@article{pan2023toward,
  title={Toward the cure of privacy policy reading phobia: Automated generation of privacy nutrition labels from privacy policies},
  author={Pan, Shidong and Hoang, Thong and Zhang, Dawen and Xing, Zhenchang and Xu, Xiwei and Lu, Qinghua and Staples, Mark},
  journal={arXiv preprint arXiv:2306.10923},
  year={2023}
}

@article{si2024solution,
  title={A Solution toward Transparent and Practical AI Regulation: Privacy Nutrition Labels for Open-source Generative AI-based Applications},
  author={Si, Meixue and Pan, Shidong and Liao, Dianshu and Sun, Xiaoyu and Tao, Zhen and Shi, Wenchang and Xing, Zhenchang},
  journal={arXiv preprint arXiv:2407.15407},
  year={2024}
}

@inproceedings{pushkarna2022data,
  title={Data cards: Purposeful and transparent dataset documentation for responsible ai},
  author={Pushkarna, Mahima and Zaldivar, Andrew and Kjartansson, Oddur},
  booktitle={Proceedings of the 2022 ACM Conference on Fairness, Accountability, and Transparency},
  pages={1776--1826},
  year={2022}
}

@article{rule_based_rewards_2023,
  title={Rule-Based Rewards for Language Model Safety},
  author={OpenAI},
  year={2023},
  url={https://cdn.openai.com/rule-based-rewards-for-language-model-safety.pdf}
}

@inproceedings{DBLP:conf/icsm/SunXX0022,
  author       = {Jiamou Sun and
                  Zhenchang Xing and
                  Xiwei Xu and
                  Liming Zhu and
                  Qinghua Lu},
  title        = {Heterogeneous Vulnerability Report Traceability Recovery by Vulnerability
                  Aspect Matching},
  booktitle    = {{IEEE} International Conference on Software Maintenance and Evolution,
                  {ICSME} 2022, Limassol, Cyprus, October 3-7, 2022},
  pages        = {175--186},
  publisher    = {{IEEE}},
  address      = {Piscataway, NJ, USA},
  year         = {2022},
  url          = {https://doi.org/10.1109/ICSME55016.2022.00024},
  doi          = {10.1109/ICSME55016.2022.00024},
  bibsource    = {dblp computer science bibliography, https://dblp.org}
}

@inproceedings{DBLP:conf/icse/SunX00023,
  author       = {Jiamou Sun and
                  Zhenchang Xing and
                  Qinghua Lu and
                  Xiwei Xu and
                  Liming Zhu},
  title        = {A Multi-faceted Vulnerability Searching Website Powered by Aspect-level
                  Vulnerability Knowledge Graph},
  booktitle    = {45th {IEEE/ACM} International Conference on Software Engineering:
                  {ICSE} 2023 Companion Proceedings, Melbourne, Australia, May 14-20,
                  2023},
  pages        = {60--63},
  publisher    = {{IEEE}},
  address      = {Piscataway, NJ, USA},
  year         = {2023},
  url          = {https://doi.org/10.1109/ICSE-Companion58688.2023.00025},
  doi          = {10.1109/ICSE-COMPANION58688.2023.00025},
  bibsource    = {dblp computer science bibliography, https://dblp.org}
}

@inproceedings{DBLP:conf/compsac/GuoXCLBZ21,
  author       = {Hao Guo and
                  Zhenchang Xing and
                  Sen Chen and
                  Xiaohong Li and
                  Yude Bai and
                  Hu Zhang},
  title        = {Key Aspects Augmentation of Vulnerability Description based on Multiple
                  Security Databases},
  booktitle    = {{IEEE} 45th Annual Computers, Software, and Applications Conference,
                  {COMPSAC} 2021, Madrid, Spain, July 12-16, 2021},
  pages        = {1020--1025},
  publisher    = {{IEEE}},
  address      = {Piscataway, NJ, USA},
  year         = {2021},
  url          = {https://doi.org/10.1109/COMPSAC51774.2021.00138},
  doi          = {10.1109/COMPSAC51774.2021.00138},
  bibsource    = {dblp computer science bibliography, https://dblp.org}
}

@inproceedings{DBLP:conf/aaai/0002WGX024,
  author       = {Shuai Guo and
                  Qiuwen Wang and
                  Yijie Gao and
                  Rong Xie and
                  Li Song},
  editor       = {Michael J. Wooldridge and
                  Jennifer G. Dy and
                  Sriraam Natarajan},
  title        = {Depth-Guided Robust and Fast Point Cloud Fusion NeRF for Sparse Input
                  Views},
  booktitle    = {Thirty-Eighth {AAAI} Conference on Artificial Intelligence, {AAAI}
                  2024, Thirty-Sixth Conference on Innovative Applications of Artificial
                  Intelligence, {IAAI} 2024, Fourteenth Symposium on Educational Advances
                  in Artificial Intelligence, {EAAI} 2024, February 20-27, 2024, Vancouver,
                  Canada},
  pages        = {1976--1984},
  publisher    = {{AAAI} Press},
  address      = {Palo Alto, California, USA},
  year         = {2024},
  url          = {https://doi.org/10.1609/aaai.v38i3.27968},
  doi          = {10.1609/AAAI.V38I3.27968},
  bibsource    = {dblp computer science bibliography, https://dblp.org}
}

@article{DBLP:journals/ieeejas/ZhuZHMWX25,
  author       = {Xiaogang Zhu and
                  Wei Zhou and
                  Qing{-}Long Han and
                  Wanlun Ma and
                  Sheng Wen and
                  Yang Xiang},
  title        = {When Software Security Meets Large Language Models: {A} Survey},
  journal      = {{IEEE} {CAA} J. Autom. Sinica},
  volume       = {12},
  number       = {2},
  pages        = {317--334},
  year         = {2025},
  url          = {https://doi.org/10.1109/JAS.2024.124971},
  doi          = {10.1109/JAS.2024.124971},
  bibsource    = {dblp computer science bibliography, https://dblp.org}
}

@article{10.1145/3745026,
author = {Yitagesu, Sofonias and Xing, Zhenchang and Zhang, Xiaowang and Feng, Zhiyong and Bi, Tingting and Han, Linyi and Li, Xiaohong},
title = {Systematic Literature Review on Software Security Vulnerability Information Extraction},
year = {2025},
publisher = {Association for Computing Machinery},
address = {New York, NY, USA},
issn = {1049-331X},
url = {https://doi.org/10.1145/3745026},
doi = {10.1145/3745026},
note = {Just Accepted},
journal = {ACM Trans. Softw. Eng. Methodol.},
month = jun
}

@inproceedings{DBLP:conf/ijcai/SkaggsRMGC24,
  author       = {Jonathan Skaggs and
                  Michael Richards and
                  Melissa Morris and
                  Michael A. Goodrich and
                  Jacob W. Crandall},
  title        = {Fostering Collective Action in Complex Societies Using Community-Based
                  Agents},
  booktitle    = {Proceedings of the Thirty-Third International Joint Conference on
                  Artificial Intelligence, {IJCAI} 2024, Jeju, South Korea, August 3-9,
                  2024},
  pages        = {211--219},
  publisher    = {ijcai.org},
  year         = {2024},
  url          = {https://www.ijcai.org/proceedings/2024/24},
  bibsource    = {dblp computer science bibliography, https://dblp.org}
}

@inproceedings{DBLP:conf/www/Zhang0SAZFS23,
  author       = {Shichang Zhang and
                  Jiani Zhang and
                  Xiang Song and
                  Soji Adeshina and
                  Da Zheng and
                  Christos Faloutsos and
                  Yizhou Sun},
  editor       = {Ying Ding and
                  Jie Tang and
                  Juan F. Sequeda and
                  Lora Aroyo and
                  Carlos Castillo and
                  Geert{-}Jan Houben},
  title        = {PaGE-Link: Path-based Graph Neural Network Explanation for Heterogeneous
                  Link Prediction},
  booktitle    = {Proceedings of the {ACM} Web Conference 2023, {WWW} 2023, Austin,
                  TX, USA, 30 April 2023 - 4 May 2023},
  pages        = {3784--3793},
  publisher    = {{ACM}},
  year         = {2023},
  url          = {https://doi.org/10.1145/3543507.3583511},
  doi          = {10.1145/3543507.3583511},
  bibsource    = {dblp computer science bibliography, https://dblp.org}
}

@article{DBLP:journals/tosem/ZhangXGXCZZFZ25,
  author       = {Sai Zhang and
                  Zhenchang Xing and
                  Ronghui Guo and
                  Fangzhou Xu and
                  Lei Chen and
                  Zhaoyuan Zhang and
                  Xiaowang Zhang and
                  Zhiyong Feng and
                  Zhiqiang Zhuang},
  title        = {Empowering Agile-Based Generative Software Development through Human-AI
                  Teamwork},
  journal      = {{ACM} Trans. Softw. Eng. Methodol.},
  volume       = {34},
  number       = {6},
  pages        = {156:1--156:46},
  year         = {2025},
  url          = {https://doi.org/10.1145/3702987},
  doi          = {10.1145/3702987},
  bibsource    = {dblp computer science bibliography, https://dblp.org}
}
